\pdfoutput=1

\documentclass[12pt]{article}
\usepackage[usenames,dvips]{color}
\usepackage[pdftex]{graphicx}
\usepackage{cite,epsfig,bbm}

\def\be{\begin{equation}}
\def\ee{\end{equation}}
\def\bea{\begin{eqnarray}}
\def\eea{\end{eqnarray}}

\def\nn{\nonumber}
\def\bbuildrel#1_#2^#3{\mathrel{\mathop{\kern 0pt#1}\limits_{#2}^{#3}}}
\def\slash#1{\setbox0=\hbox{$#1$}#1\hskip-\wd0\dimen0=5pt\advance
       \dimen0 by-\ht0\advance\dimen0 by\dp0\lower0.5\dimen0\hbox
         to\wd0{\hss\sl/\/\hss}}
\def\gev{{\rm GeV}}

\newcommand{\gae}{\lower 2pt \hbox{$\, \buildrel {\scriptstyle >}\over {\scriptstyle
\sim}\,$}}
\newcommand{\lae}{\lower 2pt \hbox{$\, \buildrel {\scriptstyle <}\over {\scriptstyle
\sim}\,$}}

\newcommand{\spp}{\vphantom{\bigg(}}

\begin{document}

\begin{titlepage}

\begin{flushright}
FERMILAB-PUB-07-315-T\\
BNL-HET-07/10\\ [2cm]
\end{flushright}

\begin{center}

\setlength {\baselineskip}{0.3in} {\bf
Footprints of the Beyond in flavor physics: \\
Possible role of the Top Two Higgs Doublet Model
}
\\[2cm]

\setlength {\baselineskip}{0.2in} {\large Enrico
Lunghi$^1$ and Amarjit Soni$^{2}$}\\[5mm]

$^1$~{\it
 Fermi National Accelerator Laboratory \\
 P.O. Box 500 , Batavia, IL, 60510-0500, USA\\
 E-mail: {\rm lunghi@fnal.gov}
}\\[5mm]
$^2$~{\it
 Physics Department, Brookhaven National Laboratory, \\
Upton, New York, 11973, USA\\
E-mail: {\rm soni@quark.phy.bnl.gov}
}\\[3cm]

\end{center}

\begin{abstract}

The B-factories results provide an impressive confirmation of the
Standard Model (SM) description of flavor and CP
violation. Nevertheless, as more data were accumulated, deviations in
the 2.5-3.5 $\sigma$ range have emerged pointing to the exciting
possibility of new CP-odd phase(s) and flavor violating parameters in
B-decays. Primarily this seems to be the case in the time dependent CP
asymmetries in penguin dominated modes (e.g. $B\to \phi (\eta')
K_s$). We discuss these and other deviations from the SM and, as an
illustration of possible new physics scenarios, we examine the role of
the Top Two Higgs Doublet Model. This is a simple extension of the SM
obtained by adding second Higgs doublet in which the Yukawa
interactions of the two Higgs doublets are assigned in order to
naturally account for the large top-quark mass. Of course, many other
extensions of the Standard Model could also account for these
experimental deviations. Clearly if one takes these deviations
seriously then some new particles in the $\approx$ 300 GeV to
$\approx$ few TeV with associated new CP-odd phase(s) are needed.

\end{abstract}

\setlength{\baselineskip}{0.2in}

\end{titlepage}

\section{Introduction}
The spectacular performance of the two asymmetric B-factories is a
triumph of accelerator science. Both machines appreciably exceeded
their designed luminosities and presently have delivered over 1
ab$^{-1}$ of data~\cite{masashi}.

On the one hand, the first crucial result is a striking confirmation
of CKM--paradigm~\cite{ckm} of flavor and CP violation. It is clear
that the CKM phase provides the dominant explanation for the observed
CP violation to an accuracy of about 10-15\%. This strongly suggests
that new physics most likely can only show up as a perturbation
requiring accurate measurements and precise theoretical calculations.

On the other hand, many B-factory results indicate interesting
deviations from the SM. One of the most compelling hints of new
physics are the measurements of the time-dependent CP asymmetries in
penguin dominated modes that turned out to be systematically smaller
than the SM expectation. While the calculation of these asymmetries
requires to keep under control long distance QCD effects, the QCD
factorization as well as several other approaches shows that some of
the modes are extremely clean ({\it i.e.}  $\phi K_s$, $\eta^\prime
K_s$ and $K_s$ $K_s$ $K_s$ final states). The magnitude of the
deviation ranges from 2.5$\sigma$ to about 4$\sigma$ depending on how
one chooses to compare. The amplitudes for these decays are dominated
by penguin ({\it i.e.} short distance) contributions: hence,
deviations in these CP asymmetries are expected and quite natural in a
very wide class of new physics scenarios. It is therefore extremely
important to follow this issue very closely.

Unfortunately a sizable reduction of the experimental errors on these
asymmetries requires significantly greater statistics and is bound
to be slow: the projected doubling of the integrated luminosity by the
end of 2008 is unlikely to resolve this issue in a decisive fashion.
From a theoretical point of view, it would be extremely desirable to
reduce the experimental uncertainties at the 5\% level because a
Standard Model irreducible pollution of a few percents is expected.
The needed luminosity for this important enterprise may have to await
the advent of a Super-B factory~\cite{
Akeroyd:2004mj,Hewett:2004tv,Albert:2005tg}.

In addition to this hint for new physics there are several other
measurements that deviate sizably from the respective SM expectations.
Among those we have the muon anomalous magnetic moment, difference in
direct CP asymmetries in $B\to K\pi$ decays and the tension between
$|V_{ub}|$ and the time-dependent CP asymmetry in $B\to J/\psi K_s$.
In this paper, we present an extensive discussion of several important
experimental hints for deviations from the SM. As an illustration we
study how a simple and well motivated extension of the SM (the Top Two
Higgs Doublet Model~\cite{Das:1995df,Kiers:1998ry,Wu:1999nc}) can
handle the experimental data.

Needless to say, the deviations seen in B decays~\cite{mn_moriond} and
some other aspects of flavor physics, may also be accountable by many
other extensions of the SM; for example supersymmetry~\cite{luca}, a
fourth family~\cite{ghou}, a Z-penguin ~\cite{gudrun}, warped
flavor-dynamics~\cite{aps} etc. Clearly, the key features of any
beyond the Standard Model scenario that is to account for the
experimental deviations in B-physics and other flavor physics that are
being discussed here are that it has to have at least one new CP-odd
phase and new particles in the $\approx$ 300 GeV to $\approx$ few TeV
range. Much more experimental information is required to disentangle
the various possibilities.

In Sec.~\ref{sec:hints} we present the list of problematic
measurements that we consider and summarize them in a {\it pull}
table. In Sec.~\ref{sec:t2hdm} we give a short overview of the Top Two
Higgs Doublet Model (T2HDM). In Sec.~\ref{sec:chisquare} we perform a
chi-squared analysis of the T2HDM and show how present experimental
results, using observables that are relatively
clean, constrain its parameter space. In Secs.~\ref{sec:charged} and
\ref{sec:neutral} we present details of the calculation of T2HDM
contributions to various observables. A brief summary and outlook is
given in Sec.~\ref{sec:summary}.

\section{Possible hints for deviations from the SM}
\label{sec:hints}
In this section we summarize some of the experimental problems that
have surfaced in the past few years connected with the Standard Model
picture of flavor physics. In particular we focus on the tension
between the measured time dependent CP asymmetry in $B\to J/\psi K_s$
and the rest of the unitarity triangle fit, the discrepancy between CP
asymmetries in $b\to s\bar s s$ ({\it e.g. }$B\to (\phi,\eta^\prime)
K_s$) and $b\to c\bar c s$ ($B\to J/\psi K_s$) transitions, the
difficulties in describing the CP asymmetries in the decays $B^0 \to
K^+ \pi^-$ and $B^- \to K^- \pi^0$, the anomalous magnetic moment of
the muon and the forward-backward asymmetry in $Z\to b\bar b$.
\subsection*{\boldmath $a_{\psi K}$: Standard Model prediction vs direct measurement}
\begin{figure}
\begin{center}
\includegraphics[width=0.48 \linewidth]{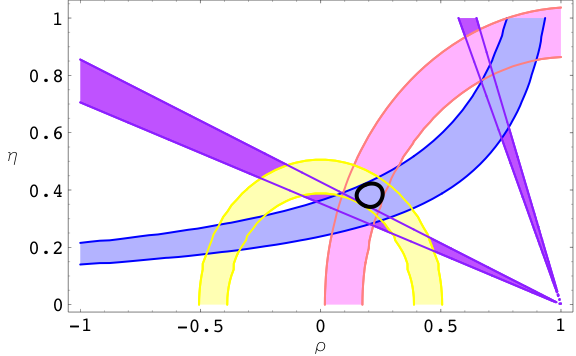}
\end{center}
\vskip -0.7cm
\caption{Unitarity triangle fit in the SM. The constraints from
$|V_{ub}/V_{cb}|$, $\varepsilon_K$, $\Delta M_{B_s}/\Delta M_{B_d}$
are included in the fit; the region allowed by $a_{\psi K}$ is
superimposed.}
\label{fig:chism}
\end{figure}

The standard unitarity triangle fit, with the inclusion of the
constraints from $|V_{ub}/V_{cb}|$, $\varepsilon_K$, $\Delta M_{B_s}$
and $\Delta M_{B_d}$ predicts $a_{\psi K} = \sin (2\beta) =0.78 \pm
0.04$. Here, we used a simple $\chi^2$ fit in which we use the inputs
given in Table~\ref{tab:ut_inputs_intro} and assume that all errors
are gaussian (this means, for instance, that we combine systematic and
statistical errors in quadrature). The direct determination of this
asymmetry via the ``gold - plated" $\psi K_s$ modes
"yields~\cite{Barberio:2006bi} $a_{\psi K}^{WA} = 0.675 \pm 0.026$ and
deviates from the SM prediction by about two standard deviations. In
Fig.~\ref{fig:chism} we show the SM fit of the unitarity triangle in
the $(\bar \rho,\bar\eta)$ plane and the $a_{\psi K}$ constraint is
superimposed. From the figure it is clear that this effect is mainly
due to the conflict between $a_{\psi K}$ and $|V_{ub}|$.  Note also
that the former measurement is clean from hadronic uncertainties and
the latter uses basically a tree--level process.

In order to test the stability of this 2$\sigma$ effect, it is useful
to entertain a scenario in which the errors on $|V_{ub}/V_{cb}|$ and
on the SU(3) breaking ratio obtained by lattice calculations $\xi_s$
are increased.  Increasing $\delta |V_{ub}/V_{cb}| = 10\%$ (from about
7\%) and $\delta \xi_s = 0.06$ (from 3-4\%), the prediction for $sin
2\beta$ does not change much: we find $\sin (2\beta) =0.78 \pm
0.05$. It is also interesting to consider the impact of the very
recent lattice determination of $\hat B_K$ presented in
Ref.~\cite{unknown:2007pb}: using $\hat B_K = 0.765 \pm 0.017 \pm
0.040$, the fit gets slightly worse and the prediction for $\sin
2\beta$ reads $0.76 \pm 0.035$. The conclusion of these exercises is
that the strain between the direct determination of $a_{\psi K}$ and
the rest of the unitarity triangle fit is quite
solid~\cite{other_fits}.
%
%%%%%%%%%%%%%%%%%%%%%%%%%%%%%%%%%%%%%%%%%%%%%%%%%%%%%%%%%%%%%%%%%%%%%%%
%%%%%%%%%%%%%%%%%%%%%%% FIGURE: ChiSM %%%%%%%%%%%%%%%%%%%%%%%%%%%%%%%%%
%%%%%%%%%%%%%%%%%%%%%%%%%%%%%%%%%%%%%%%%%%%%%%%%%%%%%%%%%%%%%%%%%%%%%%%
\subsection*{\boldmath Time--dependent CP asymmetries in $b\to s \bar s s$
modes}
Within the SM, the CP asymmetries in penguin dominated $b\to s$
transitions such as $\phi K_s$ and $\eta' K_s$ are equal to
$\sin(2\beta)$ up to penguin polluting effects, that are expected to
be fairly small in these modes~\cite{gw,Grossman:1997gr,ls}. The
calculation of matrix elements of penguin operators is an
intrinsically non--perturbative task, and it has been recently studied
using many different approaches~\cite{ccs, Beneke:2005pu,bhnr,zw}.
These studies show that while a precise calculation of hadronic
uncertainties is very difficult, at least three cases, namely $\eta'
K_s$, $\phi K_s$ and $K_s K_s K_s$~\cite{gh04}
are notably clean with only a few
percent contaminations. In many other cases rough estimates~(see for
instance Refs.~\cite{ls,Beneke:2005pu,ccs}) suggest hadronic uncertainties
to be less than 10\%. For example, Ref.~\cite{Beneke:2005pu} quotes
$a_{\eta^\prime K} - a_{\psi K} = 0.01 \pm 0.01$ and $a_{\phi K} -
a_{\psi K} = 0.02 \pm 0.01$. The measurements of the time dependent CP
asymmetries in the $\eta^\prime$ and $\phi$ modes, yield
$a_{\eta^\prime K} = 0.61 \pm 0.07$ and $a_{\phi K} = 0.39 \pm
0.18$. The latter deviates from the SM prediction $a_{s\bar s s} =
0.78\pm 0.04$ at the two sigma level.

It is rather curious that all the time dependent CP asymmetries in
$b\to s\bar s s$ have been measured to be somewhat smaller than the
$B\to J/\psi K_s$ asymmetry. If we naively compute the average of the
CP asymmetries in all the $b\to s\bar s s$ modes, even though only
three of the modes are very clean and others may have $O(10\%)$
uncertainties, one then finds $[a_{s\bar s s}]_{\rm average} = 0.52
\pm 0.05$ with a deviation of about 4$\sigma$ from the SM prediction
and about 3$\sigma$ from the value directly measured by the $\psi K_s$
method.

For the sake of completion, we also note that just averaging over the
three clean modes gives $a_{clean} = 0.57\pm 0.06$. Since in so far as
the SM is concerned, $\sin 2 \beta$ may be measured either by these
three clean penguin modes or by the $J/\psi K_s$ modes, the best
``SM'' direct measurement of $\sin 2\beta$ is given by the average over the
$(J/\psi,\phi,\eta^\prime,K_s K_s) K_s$ modes: we thus find $\sin 2 \beta
=0.66\pm0.02$ which is again about 2.5$\sigma$ from the SM prediction of
$0.78\pm 0.04$.
\begin{table}
\begin{center}
\begin{tabular}{|l|l|}
\hline
\spp  $|V_{ub}/V_{cb}| = 0.1036 \pm 0.0074$~\cite{Yao:2006px}&
      $\varepsilon_K^{\rm exp} = (2.280 \pm 0.013 ) \; 10^{-3}$ \\
\spp  $\Delta m_{B_s}^{\rm exp} = (17.77 \pm 0.10 \pm 0.07) {\rm ps}^{-1}$~\cite{Evans:2007hq}&
      $a_{\psi K_s}^{\rm exp} = 0.675 \pm 0.026$ \\
\spp  $\Delta m_{B_d}^{\rm exp} = (0.507 \pm 0.005) {\rm ps}^{-1}$ &
      $\hat B_K=0.79 \pm 0.04 \pm 0.08$~\cite{Bona:2006ah, chrisD} \\
\spp  $\xi_s = 1.210^{+0.047}_{-0.035}$~\cite{Mackenzie:2006un} &
      \\\hline
\end{tabular}
\caption{Inputs that we use in the unitarity triangle
fit.\label{tab:ut_inputs_intro}}
\end{center}
\end{table}
\subsection*{\boldmath CP asymmetries in $B\to K \pi$}
The QCD--factorization predictions for the individual CP asymmetries
in $B\to K\pi$ decays~\cite{Beneke:2000ry,Beneke:2003zv} are
extremely sensitive to non--factorizable (hence model dependent)
effects and cannot be used to directly constrain the SM. Luckily it
turns out that, in the calculation of the difference between the CP
asymmetries in $B^+\to K^+\pi^-$ and $B^- \to K^- \pi^0$, most model
dependent uncertainties cancel and the QCD--factorization prediction
is quite reliable. The magnitude of this cancelation is apparent in
the comparison between the predictions for the individual
asymmetries and for their difference. The results of
Ref.~\cite{Beneke:2003zv} read:
\bea
A_{CP} (B^-\to K^- \pi^0)      & = & \left( 7.1^{+1.7+2.0+0.8+9.0}_{-1.8-2.0-0.6-9.7} \right) \% \\
A_{CP} (\bar B^0\to K^- \pi^+) & = & \left( 4.5^{+1.1+2.2+0.5+8.7}_{-1.1-2.5-0.6-9.5} \right) \% \; ,
\eea
where the first error corresponds to uncertainties on the CKM
parameters and the other three correspond to variation of various
hadronic parameters; in particular, the fourth one corresponds to the
unknown power corrections. The main point is that the uncertainties in
the two asymmetries are highly correlated. This fact is reflected in
the prediction for their difference; we find:
\bea
\Delta A_{CP} = A_{CP} (B^-\to K^- \pi^0) - A_{CP}
(\bar B^0\to K^- \pi^+) = (2.5 \pm 1.5)\% \; .
\label{dacp_SM}
\eea
In evaluating the theory error for this case,
we followed the analysis presented in
Ref.~\cite{Beneke:2003zv} and even allowed for some extreme scenarios
(labeled S1-S4 in Ref.~\cite{Beneke:2003zv}) in which several inputs
are simultaneously pushed to the border of their allowed ranges. The
comparison of the SM prediction in Eq.~(\ref{dacp_SM}) to the
experimental determination of the same quantity~\cite{Barberio:2006bi}
\bea
\Delta A_{CP}^{\rm exp} = (14.4\pm 2.9)\% \; ,
\eea
yields a 3.5$\sigma$ effect.
\subsection*{\boldmath Muon $g-2$}
The muon anomalous magnetic moment has been thoughtfully investigated
in the literature. The most up--to--date calculation of the SM
prediction suffers from model dependent uncertainties in the
calculation of the light--by--light hadronic contribution;
nevertheless, all the estimates (see, for instance,
Ref.~\cite{Passera:2007fk} for a collection of results) point to a SM
prediction that is lower than the experimental measurement by about
three sigmas. The inconsistency between the extraction of the hadronic
contribution to the vacuum polarization from $\tau$ and $e^+ e^-$ data
is still an open question. We note, however, that the use of the
former requires model dependent assumptions on the size of isospin
breaking effects; for this reason, most analyses prefer not to include
$\tau$ decay data. The most recent theory estimate
is~\cite{Miller:2007kk}
\bea
a_\mu^{\rm SM} = 116591785 (61) \times 10^{-11} \;,
\eea
while the present measurement is~\cite{Bennett:2004pv,Bennett:2006fi}:
\bea
a_\mu^{\rm SM} = 116592080 (63) \times 10^{-11} \; .
\eea
The discrepancy is at the 3$\sigma$ level.
\subsection*{\boldmath Forward--backward asymmetry in $Z\to b\bar b$}
The LEP measurement of the forward--backward asymmetry in $Z\to b\bar b$
reads $A_{fb}^{0,b} = 0.0992 \pm 0.0016$. The discrepancy between this
experimental result and the central value of the SM fit,
$(A_{fb}^{0,b})_{SM} = 0.1038$ is at the 3 sigma level. Care has to be
taken in interpreting this result because the indirect determination
of $A_{fb}^{0,b}$ from the forward--backward Left--Right asymmetry
($A_b$) is compatible with the SM prediction at 1 sigma.
\subsection*{Overview: the pull table}
Let us give a global view of the status of the Standard Model by
collecting most measurements sensitive to the flavor sector and their
deviation from the corresponding SM predictions\footnote{See
Sec.~\ref{sec:charged} for a detailed discussion of the various
observables}. Note that several of the entries indicate deviation
from the SM in the 2.5 - 3.5 $\sigma$ range.
\begin{center}
\begin{displaymath}
\begin{tabular}{|c|c|c|c|} \hline
\spp Observable & Experiment & SM & Pull \cr
\hline\spp
${\cal B} (B\to X_s\gamma)\times 10^4$ & $3.55 \pm 0.26$ & $2.98\pm 0.26$ & +1.6 \cr
\hline\spp
${\cal B} (B\to \tau\nu_\tau)\times 10^4$ & $1.31 \pm 0.48$ & $0.85\pm 0.13$ & +0.9 \cr
\hline\spp
$\Delta m_{B_s} \; ({\rm ps}^{-1})$ & $17.77 \pm 0.12$ & $18.6\pm 2.3$ & -0.4 \cr
\hline\spp
$a_{\psi K}$ & $0.675 \pm 0.026$ & $0.78\pm 0.04$ & -2.0 \cr
\spp $a_{\phi K}$ & $0.39 \pm 0.18$ &$0.80\pm 0.04$ & -2.2 \cr
\spp $a_{\eta^\prime K}$ & $0.61 \pm 0.07$ & $0.79\pm 0.04$ & -2.0 \cr
\spp $a_{K_s K_s K_s}$ & $0.51 \pm 0.21$ & $0.80\pm 0.04$  & -1.3 \cr
\spp $a_{(\phi K + \eta^\prime K+KKK)}$ & $0.57 \pm 0.06$ &  & -2.9 \cr
\spp $a_{(\phi K + \eta^\prime K+KKK+\psi K)}$ & $0.66 \pm 0.02$ & & -2.6 \cr
\end{tabular}
\end{displaymath}
\end{center}
\begin{center}
\begin{displaymath}
\begin{tabular}{|c|c|c|c|}
\spp $[a_{s\bar s s}]_{\rm naive average}$ & $0.52 \pm 0.05$ &  &  -3.7 \cr
\hline \spp
$\Delta \Gamma_s/\Gamma_s$ & $0.27 \pm 0.08$ & $0.147\pm 0.060$ & +1.2 \cr
\hline\spp
$\Delta A_{CP}$ & $0.144 \pm 0.029$ & $0.025\pm 0.015$ & +3.6 \cr
\hline\spp
$a_\mu \times 10^{11}$ & $1.16592080(63) $ & $1.16591785(61)$ & +3.4 \cr
\hline\spp
$A_{fb}^{0,b}$ &  $0.0992 \pm 0.0016 $ & $0.1038$ & -2.9  \cr
\hline\spp
$|V_{ub}| \times 10^3$ & $4.31 \pm 0.30$ & $3.44\pm 0.16$ & +2.6 \cr
\hline
\end{tabular}
\end{displaymath}
\end{center}
\noindent
In the next section we will introduce a particular new physics model,
the two Higgs doublet model for the top quark (T2HDM), and see how
well it can accommodate the above mentioned deviations.

\section{The two Higgs doublet model for the top quark}
\label{sec:t2hdm}
The T2HDM is a special case of type-III 2HDM. It was first proposed in
Ref.~\cite{Das:1995df} and subsequently analyzed in
Refs.~\cite{Kiers:1998ry,Wu:1999nc,Xiao:2006dq}. In this model, one of
the Higgses has only interactions involving the right--hand top, while
the other one couples to the remaining right--handed fermions (but not
to the top). The main motivation for this model is to give the top
quark a unique status, thus explaining in a natural way its large
mass; hence large values of $\tan \beta_H$ (the ratio of the vev's of
the two Higgs fields) are preferred. As we will see in the following,
a consequence of the peculiar structure of the T2HDM is that the model
contains two additional flavor changing complex couplings on top of
the standard 2HDM parameters.

The Yukawa interactions of the quarks with the Higgs fields are:
\bea
{\cal L}_Y & = & - {\bar Q}_L H_1 Y_d d_R
                 - {\bar Q}_L {\tilde H}_1 Y_u {\mathbbm 1}^{(12)} u_R
                 - {\bar Q}_L {\tilde H}_2 Y_u {\mathbbm 1}^{(3)} u_R
                 + {\rm h.c.} \; ,
\eea
where $H_i$ are the two doublets, ${\tilde H}_i = i \sigma^2 H_i^*$,
$Y_{u,d}$ are Yukawa matrices, ${\mathbbm 1}^{(12)} = {\rm
diag}(1,1,0)$, and ${\mathbbm 1}^{(3)} = {\rm diag}(0,0,1)$. After the
electroweak symmetry breaking, the neutral components of Higgs
doublets receive two independent complex vev's, $v_1/\sqrt{2} = v e^{i
\phi_1}\cos\beta_H /\sqrt{2}$ and $v_2/\sqrt{2} = v e^{i \phi_2}
\sin\beta_H /\sqrt{2}$, whose ratio is $\tan \beta_H \equiv |v_2 / v_1|$.

The quark mass matrices in the mass eigenstate basis are:
\bea
m_d & = & D_L^\dagger \left( \frac{v_1^*}{ \sqrt{2}}  Y_d \right) D_R \; , \\
m_u & = & U_L^\dagger \left( \frac{v_1^*}{\sqrt{2}} Y_u {\mathbbm 1}^{(12)}
          + \frac{v_2^*}{\sqrt{2}} Y_u {\mathbbm 1}^{(3)} \right) U_R\; ,
\eea
where $m_{u,d}$ are diagonal, $U_{L,R}$ and $D_{L,R}$ are unitary
matrices and $V = U_L^\dagger D_L^{}$ is the CKM matrix.  The charged
and neutral Higgses interactions read:
\bea
{\cal L}_Y^C & = & -  {\bar u}_L V m_d d_R \frac{H_1^+}{v_1^*/\sqrt{2}}
                   -  {\bar u}_R \left(
                      m_u V \frac{H_1^+}{v_1^*/\sqrt{2}}
                      + \Sigma^\dagger V \left[\frac{H_2^+}{v_2^*/\sqrt{2}}-\frac{H_1^+}{v_1^*/\sqrt{2}} \right] \right) d_L + {\rm h.c.}\nn\\
& & \nn \\
             & = & \frac{g_2}{\sqrt{2} m_W} {\bar u} \Bigg[
                    \left( - V m_d P_R  + m_u V P_L \right)  (G^+ - \tan \beta_H H^+)
               + \Sigma^\dagger V P_L (\tan \beta_H + \cot \beta_H) H^+ \Bigg] d + {\rm h.c.} \nonumber \\
& & \nn \\
             & \equiv &
                        {\bar u}_L ( P^H_{LR} H^+ + P^G_{LR} G^+ ) d_R +
                        {\bar u}_R ( P^H_{RL} H^+ + P^G_{RL} G^+ )
                        d_L + {\rm h.c.}   \label{Lcharged} \\
& & \nn \\
{\cal L}_Y^N & = & -  {\bar d}_L m_d d_R
\frac{H_1^{0*}}{v_1^*/\sqrt{2}}
                   -  {\bar u}_L \left(
                      m_u \frac{H_1^{0*}}{v_1^*/\sqrt{2}}
                      + \Sigma \left[\frac{H_2^{0*}}{v_2^*/\sqrt{2}}-\frac{H_1^{0*}}{v_1^*/\sqrt{2}} \right] \right) u_R +{\rm h.c.} \nn \\
& & \nn \\
             & = & \frac{g_2 \tan\beta_H}{2 m_W} \Bigg[
                   \left( {\bar d}_L m_d d_R + {\bar u}_L m_u  u_R \right)
                     \frac{h^0 \sin\alpha_H  - H^0 \cos\alpha_H}{\sin\beta_H}
                     + i  \left( {\bar d}_L m_d \gamma_5 d_R + {\bar u}_L m_u \gamma_5  u_R \right) A^0 \nn \\
& & \nn \\
             & &     - {\bar u}_L \Sigma^\dagger u_R
                \frac{H^0 \sin(\alpha_H-\beta_H) + h^0 \cos (\alpha_H-\beta_H) }{\sin^2 \beta_H}
                - i (1 + \cot^2 \beta_H) {\bar u}_L \Sigma^\dagger \gamma_5 u_R A^0
              \Bigg] + {\rm h.c.} \nn \\
& & \nn \\
             & \equiv &
                        {\bar d}_L ( P^{h^0}_d h^0 + P^{H^0}_d H^0 + i \gamma_5 P^{A^0}_d A^0  ) d_R +
                        {\bar u}_L ( P^{h^0}_u h^0 + P^{H^0}_u H^0 + i \gamma_5 P^{A^0}_u A^0  ) u_R
                        + {\rm h.c.}   \; ,\label{Lneutral}
\eea
where $\Sigma \equiv m_u U_R^\dagger {\mathbbm 1}^{(3)} U_R$. The
would be Goldstone boson $G^\pm$, the charged Higgs $H^\pm$, the heavy
and light scalars $H^0$ and $h^0$, and the pseudoscalar $A^0$ are
given by:
\bea
\pmatrix{H_1^0 e^{- i \phi_1} \cr H_2^0  e^{- i  \phi_2}} & = & \frac{1}{\sqrt{2}} \left[
         R_{\alpha_H} \pmatrix{H^0\cr h^0 \cr} + i R_{\beta_H} \pmatrix{G^0\cr A^0 \cr} +  \pmatrix{|v_1|\cr |v_2| \cr} \right] \; , \\
\pmatrix{H_1^\pm \cr H_2^\pm \cr } & = &
         R_{\beta_H} \pmatrix{G^\pm\cr H^\pm \cr} \; ,
\eea
with
\bea
R_\omega & = & \pmatrix{ \cos\omega & - \sin \omega \cr \sin \omega & \cos \omega \cr} \;
\eea
The explicit expressions for the charged Higgs couplings are:
\bea
P_{LR}^H & = & \frac{g_2}{\sqrt{2} m_W} \tan\beta_H \; V \; m_d \; ,\label{phlr}\\
P_{RL}^H & = & \frac{g_2}{\sqrt{2} m_W} \tan\beta_H \left[ (1 + \tan^{-2} \beta_H)
                                \Sigma^\dagger - m_u \right]  V \; ,\label{phrl}\\
P_{LR}^G & = & - \frac{g_2}{\sqrt{2} m_W} V\; m_d \; , \\
P_{RL}^G & = & \frac{g_2}{\sqrt{2} m_W} m_u \; V \; .
\eea
The explicit expressions for the neutral Higgs couplings are:
\bea
P_{u}^{h^0} & = & \frac{g_2 \tan\beta_H}{2 m_W} \left( m_u \frac{\sin\alpha_H}{\sin\beta_H} -
                              \Sigma^\dagger \frac{\cos(\alpha_H-\beta_H)}{\sin^2\beta_H} \right) \\
P_{u}^{H^0} & = & - \frac{g_2 \tan\beta_H}{2 m_W} \left( m_u \frac{\cos\alpha_H}{\sin\beta_H}
                              + \Sigma^\dagger \frac{\sin(\alpha_H-\beta_H)}{\sin^2\beta_H} \right) \\
P_{u}^{A^0} & = & \frac{g_2 \tan\beta_H}{2 m_W} \left( m_u
                              - \frac{1+\tan^2\beta_H}{\tan^2\beta_H} \Sigma^\dagger \right) \\
P_{d}^{h^0} & = & \frac{g_2 \tan\beta_H}{2 m_W}  m_d \frac{\sin\alpha_H}{\sin\beta_H}  \\
P_{d}^{H^0} & = & - \frac{g_2 \tan\beta_H}{2 m_W}  m_d \frac{\cos\alpha_H}{\sin\beta_H} \\
P_{d}^{A^0} & = & \frac{g_2 \tan\beta_H}{2 m_W} m_d   \; .
\eea
From the definition of $\Sigma$ it is clear that only the third row of
the matrix $U_R$ is relevant up to an overall phase (i.e. we can take
$(U_R)_{33}$ real). Taking into account the unitarity constraint, it
follows that $\Sigma$ depends on only 4 real parameters. This
statement can be explicitly verified by employing the most general
parametrization of a unitary matrix: $U = P_1 V P_2$, where $P_i$ are
diagonal phase matrices and $V$ is a unitary matrix that depends on
three angles and a single phase (e.g. it is CKM-like). The third row
of this matrix can always be written as:
\bea
U_R & = & \pmatrix{ * & * & * \cr * & * & * \cr
\hat \xi' \sqrt{1-|\hat \xi|^2} &
\hat \xi &
\sqrt{1-|\hat \xi|^2}\sqrt{1-|\hat \xi'|^2} \cr }
\eea
where $\hat \xi$ and $\hat \xi'$ are complex parameters with
$|\hat\xi^{(')}| \leq 1$. In models based on dynamical
top-condensation~\cite{Bardeen:1989ds,Miransky:1988xi} and
top-color~\cite{Hill:1991at,Buchalla:1995dp} the parameters $\hat
\xi^{(\prime)}$ are naturally of order $\epsilon_{ct} = m_c/m_t$ (see
also Ref.~\cite{Cheng:1987rs}); for this reason we introduce new
parameters $\xi^{(\prime)} = \epsilon_{ct} \hat \xi^{(\prime)}$ with
$\xi^{(\prime)} = O(1)$.  Neglecting terms proportional to the u-quark
mass, the matrix $\Sigma$ reads:
\bea
\frac{\Sigma}{m_t}
& = &
\pmatrix{
0 & 0 & 0 \cr
\epsilon_{ct}^3 \xi^* \xi^\prime \sqrt{1-|\hat\xi|^2} &
\epsilon_{ct}^3 |\xi|^2 &
\epsilon_{ct}^2 \xi^* \sqrt{1-|\hat \xi|^2}\sqrt{1-|\hat \xi^\prime|^2} \cr
\epsilon_{ct} \xi^\prime \sqrt{1-|\hat \xi^\prime|^2} (1-|\hat \xi|^2) &
\epsilon_{ct} \xi \sqrt{1-|\hat \xi^\prime|^2} \sqrt{1-|\hat \xi^\prime|^2} &
(1-|\hat \xi|^2)(1-|\hat \xi^\prime|^2) \cr} \nonumber \\
& = &
\pmatrix{
0 & 0 & 0 \cr
0  & 0 & \epsilon_{ct}^2 \xi^*  \cr
\epsilon_{ct} \xi^\prime  & \epsilon_{ct} \xi  & 1 \cr
} \times  + O\left(\epsilon_{ct}^3 , \frac{m_u}{m_t} \right) \; .
\label{sigma}
\eea
From Eq.~(\ref{Lcharged}) we find the following charged Higgs
interactions between right-handed up quarks and left-handed down
quarks:
\bea
\frac{g_2 m_c \tan\beta_H}{\sqrt{2} m_W} \pmatrix{
\xi^{\prime *} \; V_{td}  &
\xi^{\prime *} \; V_{ts}  &
\xi^{\prime *} \; V_{tb}  \cr
\xi^* \; V_{td} - V_{cd} &
\xi^* \; V_{ts} - V_{cs} &
\xi^* \; V_{tb} - V_{cb} \cr
V_{td} \cot^2 \beta_H/\epsilon_{ct} + \epsilon_{ct} \xi V_{cd}  &
V_{ts} \cot^2 \beta_H/\epsilon_{ct} + \epsilon_{ct} \xi V_{cs}  &
V_{tb} \cot^2 \beta_H/\epsilon_{ct}  \cr
} \; .
\eea
In particular $\bar t_R q_L$ (q=d,s) interactions are dominated by the
$\xi$ term for $\tan\beta_H > 10$. In conclusion, the parameters of the
models are: $\tan\beta_H$, $\alpha_H$, $m_{H^\pm}$, $m_{H^0}$, $m_{h^0}$,
$m_A^0$, $\xi$ and $\xi^\prime$.

Finally let us comment on the renormalization scheme of the quark
masses that appear in Eqs.~(\ref{Lcharged}) and (\ref{Lneutral}).  In
the calculation of the additional matching conditions to various
operators we integrate out the charged and neutral higgses at some
high scale $\mu_0 \sim O (m_W, m_t)$; therefore, it is most natural to
evaluate all the relevant couplings in the $\overline {MS}$ scheme at
the high scale. This observation has a very strong impact on the
phenomenology of the T2HDM because of the strong renormalization scale
dependence of the charm quark: $m_c^{\overline {MS}}(\mu_0)/ m_c^{\rm
pole} \simeq 0.45$.

\section{Global analysis}
\label{sec:chisquare}
In this section we present the results of global $\chi^2$ fit of the
T2HDM parameter space. Here we just focus on the outcome of the fit
and investigate how well the T2HDM can answer to the problems we
collected in Sec.~\ref{sec:hints}. A detailed discussion of the
various observables that we consider is given in
Secs.~\ref{sec:charged} and \ref{sec:neutral}), in which we collect
the experimental data and the analytic formulae required to calculate
T2HDM effects. In those sections we also show the separate impact of
each observable on the T2HDM parameter space.

%%%%%%%%%%%%%%%%%%%%%%%%%%%%%%%%%%%%%%%%%%%%%%%%%%%%%%%%%%%%%%%%%%%%%%%
%%%%%%%%%%%%%%%%%%%%%%% FIGURE: Chi1 %%%%%%%%%%%%%%%%%%%%%%%%%%%%%%%%%%
%%%%%%%%%%%%%%%%%%%%%%%%%%%%%%%%%%%%%%%%%%%%%%%%%%%%%%%%%%%%%%%%%%%%%%%
\begin{figure}[t]
\begin{center}
\includegraphics[width=0.3 \linewidth]{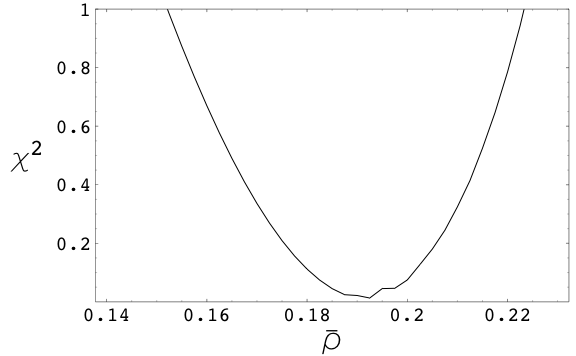}
\raisebox{0.0cm}{
\includegraphics[width=0.3 \linewidth]{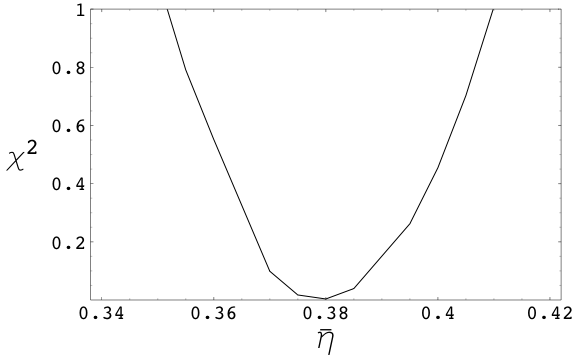}
}
\includegraphics[width=0.3 \linewidth]{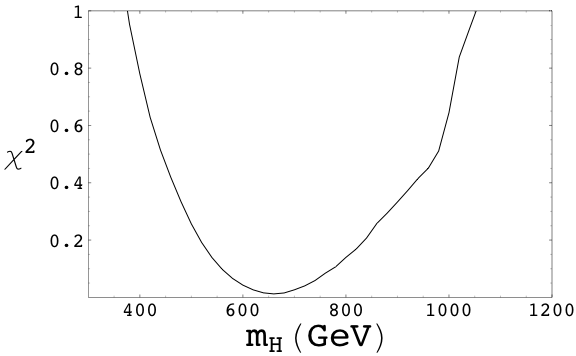}
\raisebox{0.0cm}{
\includegraphics[width=0.3 \linewidth]{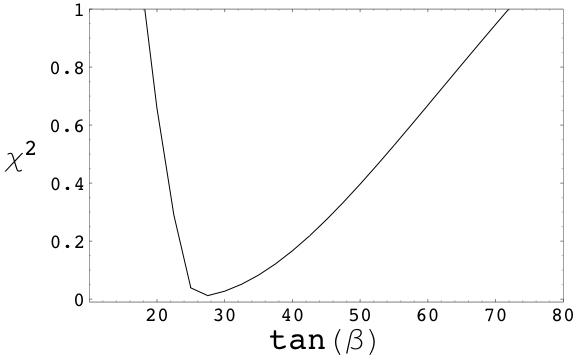}
}
\includegraphics[width=0.3 \linewidth]{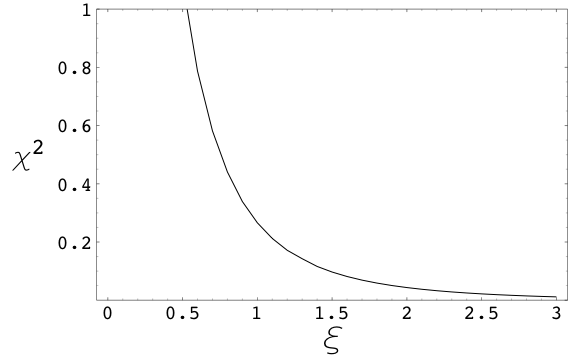}
\raisebox{0.0cm}{
\includegraphics[width=0.3 \linewidth]{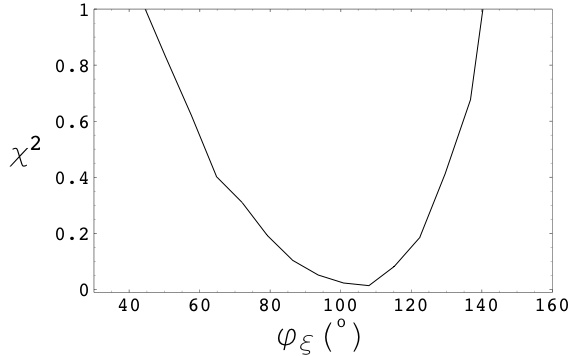}
}
\includegraphics[width=0.3 \linewidth]{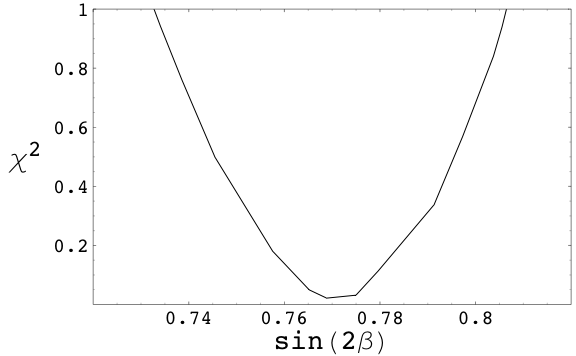}
\raisebox{0.0cm}{
\includegraphics[width=0.3 \linewidth]{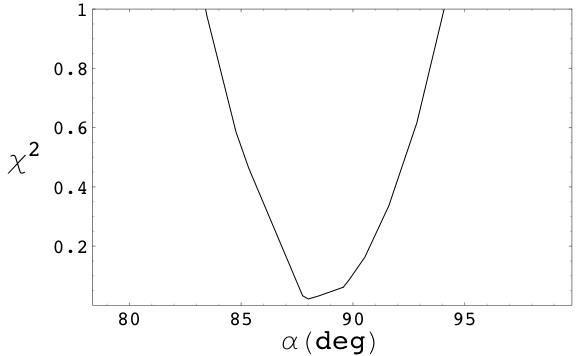}
}
\includegraphics[width=0.3 \linewidth]{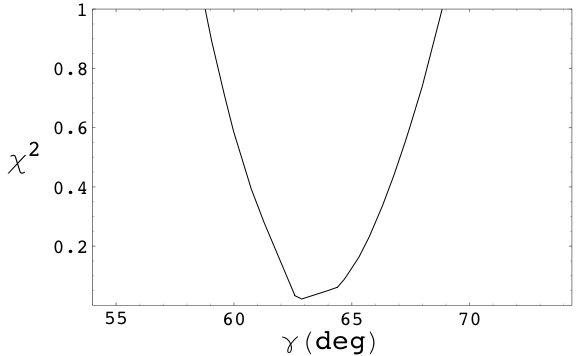}
\end{center}
\vskip -0.7cm
\caption{ Dependence of $\chi^2_{min}$ on $\bar \rho$, $\bar \eta$,
$m_{H^\pm}$, $\tan\beta_H$, $|\xi|$, $\varphi_\xi$, $\sin 2\beta$,
$\alpha$ and $\gamma$. For each value of the parameter on the x-axis,
we minimize the chi-square with respect to all the others (including
$\bar\rho$ and $\bar\eta$.}
\label{fig:chi1}
\end{figure}
%%%%%%%%%%%%%%%%%%%%%%%%%%%%%%%%%%%%%%%%%%%%%%%%%%%%%%%%%%%%%%%%%%%%%%%
We classify the various observables we consider according to whether
neutral Higgs exchange contributions are relevant or not. In the
latter case, the parameter count of the model is reduced to the sole
$\tan\beta_H$, $m_{H^\pm}$, $\xi$ and $\xi^\prime$. Observables
insensitive to the neutral Higgs sector of the T2HDM include: rare B
decays ($b\to s \gamma$, $b\to s\ell^+ \ell^-$, $B\to \tau\nu$),
neutral meson mixing ($K$, $B_d$, $B_s$, $D$), various CP asymmetries
(time--dependent asymmetries in $b\to c\bar c s$ and $b\to s\bar s s$
decays, asymmetries in flavor specific B decays, direct asymmetries in
the $B\to K\pi$ system) and the neutron electric dipole moment
(EDM). Among those observables that display some sensitivity to the
neutral Higgs sector we consider the muon anomalous magnetic moment,
$\Delta\rho$ and the $Z\to b\bar b$ vertex ($R_b$ and the
forward--backward asymmetry $A_b$).

In the $\chi^2$ analysis we focus on the first set of observables and
treat separately the $\xi^\prime=0$ and $\xi^\prime \neq 0$ cases. In
fact, the parameter $\xi^\prime$ is related exclusively to transitions
between the first and third generations and impacts only $B\to \tau
\nu$, $D\bar D$ mixing and the neutron EDM, while being completely
negligible in all other observables. The T2HDM phenomenology of
observables dominated by neutral Higgs exchanges is very similar to
the one of a regular Two Higgs Doublet Model and we will briefly
summarize it in Sec.~\ref{sec:neutral}.

Our general strategy is to include directly into the fit only
processes for which the theory error is reasonably under control; once
a region of the T2HDM parameter space has been singled out, we look at
the other observables.

\subsection*{\boldmath $\xi^\prime = 0$}
As a first step we set $\xi^\prime=0$. The $\chi^2$ that we consider
includes the following quantities: $|V_{ub}/V_{cb}|$, $\Delta
M_{B_s}/\Delta M_{B_d}$, $a_{\psi K}$, $\varepsilon_K$, $B\to X_s
\gamma$, $B \to \tau \nu$. The resulting function depends on the CKM
parameters $\bar \rho$ and $\bar \eta$, and on the T2HDM parameters
$m_{H^\pm}$, $\tan\beta_H$, $\xi = |\xi| e^{i \varphi_\xi}$.

%%%%%%%%%%%%%%%%%%%%%%%%%%%%%%%%%%%%%%%%%%%%%%%%%%%%%%%%%%%%%%%%%%%%%%%
%%%%%%%%%%%%%%%%%%%%%%% FIGURE: ChiXI %%%%%%%%%%%%%%%%%%%%%%%%%%%%%%%%%
%%%%%%%%%%%%%%%%%%%%%%%%%%%%%%%%%%%%%%%%%%%%%%%%%%%%%%%%%%%%%%%%%%%%%%%
\begin{figure}
\begin{center}
\includegraphics[width=0.48 \linewidth]{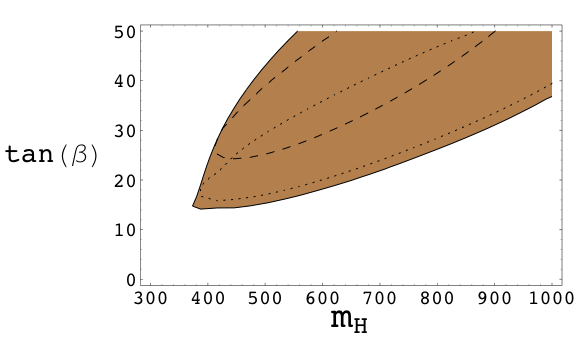}
\raisebox{0.1cm}{\includegraphics[width=0.46 \linewidth]{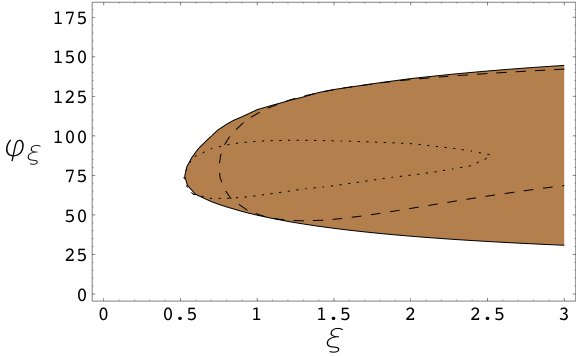}}
\end{center}
\vskip -0.7cm
\caption{ Contour plots corresponding to $\chi^2_{min} \leq 1$ in the
$(m_{H^\pm},\tan\beta_H)$ and $(\xi,\varphi_\xi)$ planes. For each point
on the contour, we minimize with respect to all other variables. The
dashed and dotted contours correspond to $\xi=(1,2)$ and
$\tan\beta_H=(30,50)$ for the left and right plot, respectively.}
\label{fig:chixi}
\end{figure}

Note that, without the inclusion of T2HDM contributions, the overall
$\chi^2$--fit in the SM is relatively poor ($\chi^2_{min} \sim
6$). Once T2HDM effects are included, the fit improves drastically and
we find $\chi^2_{min} \sim 0$. This implies that this set of
measurements singles out a clear sector of the parameter space not
compatible with the T2HDM decoupling limit. In Fig.~\ref{fig:chism} we
show the unitarity triangle fit in the Standard Model; note, in
particular, the tension between the black contour and the constraint
from $a_{\psi K}$ (not included in the fit). In Fig.~\ref{fig:chi1},
we show the actual dependence of the full $\chi^2$ on the CKM angles
and the four T2HDM parameters. The 68\% C.L. intervals that we find
are:
\bea
m_{H^\pm} &=& \left( 660^{+390}_{-280} \right) \; \gev \;,\\
\tan\beta &=& 28^{+44}_{-8} \;, \\
\xi &>& 0.5  \;,\\
\varphi_\xi & = & \left( 110^{+30}_{-65} \right)^o \;,\\
\bar \rho & = & 0.19 \pm 0.035 \;,\\
\bar \eta & = & 0.38 \pm 0.03 \;.
\eea
The corresponding ranges for the three UT angles are:
\bea
\sin (2\beta) &=& 0.77\pm0.04 \; ,\\
\alpha &=& (89\pm 6)^o \; ,\\
\gamma &=& (64\pm 5)^o \; .
\eea
In Fig.~\ref{fig:chixi} we show the correlation between these
parameters; the shaded areas correspond to $\chi^2_{min} \leq 1$ and
their projections on the axes yield the corresponding $1\sigma$
regions.

%%%%%%%%%%%%%%%%%%%%%%%%%%%%%%%%%%%%%%%%%%%%%%%%%%%%%%%%%%%%%%%%%%%%%%%
%%%%%%%%%%%%%%%%%%%%%%% FIGURE: ChiXIP %%%%%%%%%%%%%%%%%%%%%%%%%%%%%%%%
%%%%%%%%%%%%%%%%%%%%%%%%%%%%%%%%%%%%%%%%%%%%%%%%%%%%%%%%%%%%%%%%%%%%%%%
\begin{figure}
\begin{center}
\includegraphics[width=0.48 \linewidth]{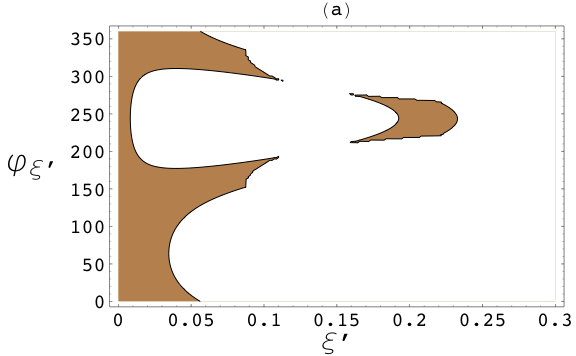}
\raisebox{0.1cm}{\includegraphics[width=0.46 \linewidth]{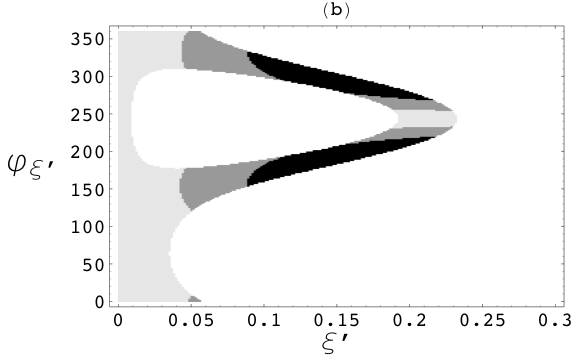}}
\end{center}
\vskip -0.7cm
\caption{Contour plot corresponding to $\chi^2_{min} \leq 1$ in the
$(\xi^\prime,\varphi_{\xi^\prime})$ plane. The rest of the parameters
are chosen so to minimize the $\chi^2$ for $\xi^\prime=0$. In the plot
on the right the light gray, dark gray and black regions correspond to
a neutron--EDM (given in units of $10^{-26}\; e \; {\rm cm}$) smaller
than 3, between 3 and 6.3, and bigger than 6.3, respectively.}
\label{fig:chixip}
\end{figure}
We are now in the position of evaluating how well the T2HDM does with
respect to the pull table we introduced in Sec.~\ref{sec:hints}. From
the outcome of the fit it is clear that the T2HDM can easily
accommodate the deviations in $B\to X_s \gamma$, $B\to \tau\nu$,
$a_{\psi K}$ and $|V_{ub}|$. Unfortunately, for $m_{H^\pm} > 400 \;
{\rm GeV}$, it seems quite difficult to accommodate the effect required
to reconcile the CP asymmetries in $B\to (\eta^\prime,\phi) K_S$ with
experimental data (see Fig.~\ref{fig:apsiks} in
Sec.~\ref{sec:apsiks}). The impact of the T2HDM on $\Delta M_{B_s}$ is
not very large and is perfectly compatible with the present
experimental determination. Finally we do not find any large
contribution to the $CP$ asymmetries in $B\to K\pi$, hence within the
T2HDM the 3.6$\sigma$ observed deviation remains unexplained.

Let us note that the solution of the $|V_{ub}|$ vs $a_{\psi K}$ puzzle
is achieved via sizable and highly correlated contributions to both
$a_{\psi K}$ and $\varepsilon_K$: after the inclusion of the
constraints from $B\to X_s \gamma$ and $B\to \tau\nu$, the solution of
this puzzle was a crucial bonus that we could not enforce (due to the
extremely reduced number of parameters that we are considering).

Finally we point out that T2HDM effects on $a_{\psi K}$ are caused by
large complex contributions to the amplitude ${\cal A} (B\to J/\psi
K_s)$ and not to the $B-\bar B$ mixing matrix element
(i.e. $M_{12}^{d}$). Since the former is dominated by the tree--level
transition $b\to c \bar c s$, any other process controlled by this
quark--level decay will display similar large effects. This is
particularly true for time dependent CP asymmetries in $B_s$ decays.
The $B_s\to J/\psi \; \eta^\prime$ mode, for instance, is based on the
$b\to c\bar c s$ amplitude, hence, in the naive factorization limit,
the T2HDM contributions to its time dependent CP asymmetry must be
identical to the corresponding ones in $B\to J/\psi K_s$. Therefore,
the above $\chi^2$ analysis predicts the T2HDM contribution to this
asymmetry to be in the $+10\%$ range. Given that the SM expectation
for this quantity is extremely small (the phase of the SM $B_s-\bar
B_s$ amplitude is about one degree), the measurement of a large
enhancement in the $B\to J/\psi K$ asymmetry is a clear indication for
a resolution of the $a_{psi K}$ puzzle via new physics in the
amplitudes (as it is the case in the T2HDM).

\subsection*{\boldmath $\xi^\prime \neq 0$}
In order to study the effects of non vanishing $\xi^\prime$, we fix
the other parameters to the values that minimize the $\chi^2$ we just
studied; then we include contributions from $B\to \tau \nu$, $D\bar D$
mixing and the neutron EDM (for the latter two, we impose upper limits
-- see Secs.~\ref{subsec:btaunu} and \ref{subsec:nEDM} for
details). Note that without the inclusion of $B\to \tau \nu$, the fit
for $\xi^\prime = 0$ favors values of $\xi$ smaller than 1 (the actual
value that we use in the $\xi^\prime \neq 0$ fit is $\xi \simeq 0.8$).

In Fig.~\ref{fig:chixip}, we plot the region of $(\xi^\prime,
\varphi_{\xi^\prime})$ plane for which this new $\chi^2$ is smaller
than 1. The main constraint comes from $B\to\tau\nu$, whose branching
ratio is proportional to $\xi^{\prime 2}$. It is interesting to
dissect contributions to the neutron EDM: in the right plot in
Fig.~\ref{fig:chixip} the regions with increasing darkness correspond
to a neutron-EDM (in units of $10^{-26}\; e \; {\rm cm}$) smaller than
3, between 3 and 6.3, and bigger than 6.3, respectively.

\section{Observables: the charged Higgs sector}
\label{sec:charged}
\subsection{$B\to X_s\gamma$ and $B\to X_s \ell^+ \ell^-$}
\label{sec:bsgamma}

%%%%%%%%%%%%%%%%%%%%%%%%%%%%%%%%%%%%%%%%%%%%%%%%%%%%%%%%%%%%%%%%%%%%%%%
%%%%%%%%%%%%%%%%%%%%%%% FIGURE: BSGAMMA %%%%%%%%%%%%%%%%%%%%%%%%%%%%%%%
%%%%%%%%%%%%%%%%%%%%%%%%%%%%%%%%%%%%%%%%%%%%%%%%%%%%%%%%%%%%%%%%%%%%%%%
\begin{figure}
\begin{center}
\includegraphics[width=0.48 \linewidth]{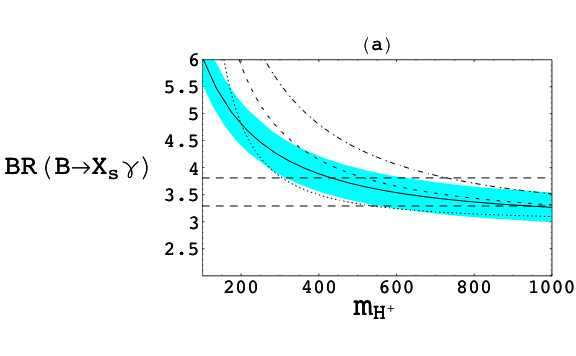}
\raisebox{0.25cm}{\includegraphics[width=0.44 \linewidth]{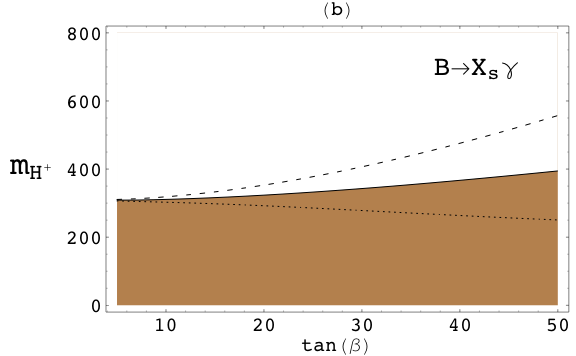}}
\end{center}
\vskip -1cm
\caption{
{\bf Plot a}. $m_{H^\pm}$ dependence of the branching ratio $B\to X_s
\gamma$ in units of $10^{-4}$. Solid, dashed, dotted and dotted-dashed
lines correspond to $(\tan\beta_H,\xi)=(10,0)$, $(50,0)$, $(50,1)$ and
$(50,-1)$, respectively. There is no appreciable dependence on
$\xi^\prime$. The two horizontal dashed lines are the experimental
68\%C.L. allowed region. The blue region represents the theory
uncertainty associated to the solid line (similar bands can be drown
for the other cases).
{\bf Plot b}. Portion of the $(\tan\beta_H,m_{H^\pm})$ plane excluded at
68\%C.L. by the $B\to X_s \gamma$ measurement. The shaded area
corresponds to $\xi=0$. The dotted and dashed lines show how this
region changes for $\xi=1$ and $-1$, respectively.
}
\label{fig:bsg}
\end{figure}

The experimental world average from the CLEO~\cite{cleobsg},
Belle~\cite{bellebsg1,bellebsg2} and
BaBar~\cite{babarbsg1,babarbsg2} collaborations is given
by~\cite{hfag}:
\begin{eqnarray}
{\rm BR} (B\to X_s \gamma)_{E_\gamma > 1.6 {\rm GeV}} = ( 3.55 \pm
0.24_{-0.10}^{+0.09}\pm 0.03) \times 10^{-4}\; .
\end{eqnarray}
The $B\to X_s \ell^+ \ell^-$ branching ratio has been recently
measured by both Belle~\cite{Iwasaki:2005sy} and
BaBar~\cite{Aubert:2004it}; in the low dilepton invariant mass
region, $1\;\gev^2 < m^2_{\ell\ell} < 6\;\gev^2$, the experimental
results read
\bea {\cal B} (B\to X_s \ell^+\ell^-) &=& (1.493 \pm
0.504^{+0.411}_{-0.321})
\times 10^{-6} \;\;\; ({\rm Belle}) \; ,\\
{\cal B} (B\to X_s \ell^+\ell^-) &=& (1.8 \pm 0.7\pm0.5) \times
10^{-6} \;\;\; ({\rm BaBar}) \; . \eea
This leads to a world average
\bea {\cal B} (B\to X_s \ell^+ \ell^-) &=& (1.60 \pm 0.51)\times
10^{-6} \; . \eea
The effective Hamiltonian responsible for the transitions $b\to s
\gamma$ and $ b\to s \ell^+ \ell^-$ is~\cite{Huber:2005ig}
\bea {\cal H}_{\rm eff} & = & - 4 \frac{G_F}{\sqrt{2}} V_{tb}^{}
V_{ts}^{*} \Bigg[ \sum_{i=1}^{10} C_i (\mu) P_i (\mu) +
\sum_{i=3}^{6} C_{iQ}(\mu) P_{iQ} + C_b(\mu) P_b \Bigg] \eea
where the most relevant operators are
\bea
P_7 &  = &  \frac{e}{16 \pi^2} m_b (\bar{s}_L \sigma^{\mu \nu}     b_R) F_{\mu \nu} \; , \\
P_8  & = &  \frac{g}{16 \pi^2} m_b (\bar{s}_L \sigma^{\mu \nu} T^a b_R) G_{\mu \nu}^a \; ,  \\
P_9   &   = & (\bar{s}_L \gamma_{\mu} b_L) \sum_\ell (\bar{\ell}\gamma^{\mu} \ell) \; , \\
P_{10} &  = & (\bar{s}_L \gamma_{\mu}     b_L) \sum_\ell (\bar{\ell}\gamma^{\mu} \gamma_5 \ell) \; .
\eea

%%%%%%%%%%%%%%%%%%%%%%%%%%%%%%%%%%%%%%%%%%%%%%%%%%%%%%%%%%%%%%%%%%%%%%%
%%%%%%%%%%%%%%%%%%%%%%% FIGURE: BSMUMU %%%%%%%%%%%%%%%%%%%%%%%%%%%%%%%%
%%%%%%%%%%%%%%%%%%%%%%%%%%%%%%%%%%%%%%%%%%%%%%%%%%%%%%%%%%%%%%%%%%%%%%%
\begin{figure}
\begin{center}
\includegraphics[width=0.48 \linewidth]{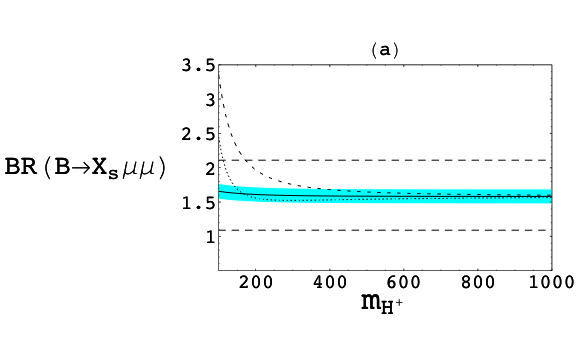}
\raisebox{0.25cm}{\includegraphics[width=0.44 \linewidth]{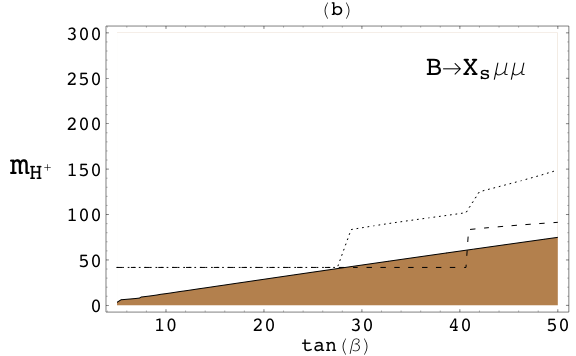}}
\end{center}
\vskip -1cm
\caption{
{\bf Plot a}. $m_{H^\pm}$ dependence of the branching ratio $B\to X_s
\mu\mu$ in units of $10^{-6}$. Solid, dashed and dotted lines
correspond to $(\tan\beta_H,\xi)=(10,0)$, $(50,1)$ and $(50,-1)$,
respectively. There is no appreciable dependence on $\xi^\prime$. The
two horizontal dashed lines are the experimental 68\%C.L. allowed
region. The blue region represents the theory uncertainty associated
to the solid line (similar bands can be drown for the other cases).
{\bf Plot b}. Portion of the $(\tan\beta_H,m_{H^\pm})$ plane excluded at
68\%C.L. by the $B\to X_s \mu\mu$ measurement. The shaded area
corresponds to $\xi=0$. The dotted and dashed lines show how this
region changes for $\xi=1$ and $-1$, respectively.
}
\label{fig:bsee}
\end{figure}

The leading order charged Higgs contributions in the T2HDM to the
Wilson coefficients $C_{7,8,9,10}$ have been in explicitly calculated
in Refs.~\cite{Kiers:1998ry,Wu:1999nc,Kiers:2000xy,Xiao:2006dq} (see
Eqs.~(7-15) of Ref.~\cite{Xiao:2006dq}). The formula for the new
physics contribution to $C_7$ is:
\bea
C_7^{\rm NP} (m_W)
& = & \left( - V_{tb}^{} V_{ts}^* \frac{4 G_F}{\sqrt{2}} \right)^{-1}
\sum_{i=u,c,t}
\Bigg\{
\frac{(P_{LR}^H)_{i3}^{} (P_{RL}^H)_{i2}^*}{m_b \; m_{u_i}} B(y_i) +
\frac{(P_{RL}^H)_{i3}^{} (P_{RL}^H)_{i2}^*}{m_{u_i}^2} \frac{A(y_i)}{6}
\Bigg\}
\nn \\
& \simeq & - \left[ B(y_t) + \tan^2 \beta_H \; B(y_c) \right]
+ \xi^* \tan^2\beta_H
\left[ -\frac{1}{6}  \frac{V_{tb}^{}}{V_{cb}^{}} A(y_c)- \epsilon_{ct}^2  \frac{V_{cs}^{}}{V_{ts}^{}} B(y_t) \right]
\; ,
\eea
where both quantities in square brackets are positive for any choice
of $\tan\beta_H$ and $m_{H^\pm}$, $y_a = m_a^2/m_{H^\pm}^2$ and the
loop-functions $A$ and $B$ are given in Ref.~\cite{Xiao:2006dq}.

A numerical formula for the calculation of the $B\to X_s \gamma$
branching ratio is given in Ref.~\cite{Hurth:2003dk,Lunghi:2006}:
\bea \label{bsgamma}
{\cal B} (\bar B \to X_s \gamma)_{E_\gamma>1.6\; {\rm GeV}}^{\rm th}
& = &
 10^{-4} \Bigg[
2.98 + 4.743 \, |\delta C_7|^2
+ 0.789 \, |\delta C_8|^2
+{\rm Re} \Big(
\nonumber \\ & & \hskip -5cm
(-7.184 + 0.612 \; i) \; \delta C_7
+ (-2.225 -0.557 \; i) \; \delta C_8
+ (2.454 - 0.884 \; i) \;  \delta C_8 \; \delta C_7^{*}
\Big)
\Bigg] \, ,
\eea
where the leading Wilson coefficients at the high scale are given by
$C_i^{(0)}(\mu_0) = C_{i,SM}^{(0)}(\mu_0) + \delta C_i$ and the
next-to-leading matching conditions are assumed not to receive any new
physics contribution, $C_i^{(1)}(\mu_0) = C_{i,SM}^{(1)}(\mu_0)$. The
formula above has been obtain by observing that using the same
numerical inputs of Ref.~\cite{Misiak:2006zs,Misiak:2006ab} and taking
$(\mu_c,\mu_b,\,mu_0) = (1.5,2.5,120)\; {\rm GeV}$, the NLO central
value of the branching ratio coincides with the NNLO
one. Eq.~(\ref{bsgamma}) also include an estimate of the new class of
power corrections identified in Ref.~\cite{Lee:2006wn} and of the
analysis of the photon energy spectrum presented in
Ref.~\cite{Becher:2006pu}. The analyses in
Refs.~\cite{Misiak:2006zs,Becher:2006pu} yield ${\cal B}(B\to X_s
\gamma)= (2.98 \pm 0.26) \times 10^{-4}$; we will therefore assign a
theoretical error of 8.7\% to the central values calculated in
Eq.~(\ref{bsgamma}).

The Standard Model matching conditions and numerical formulae for the
calculation of the integrated $B\to X_s \ell^+ \ell^-$ branching
ratios is given in Ref.~\cite{Huber:2005ig}:
\begin{eqnarray}
\label{bsll} \hskip -3cm {\cal B}_{\ell\ell}
& = &
\Big[\;2.1913 -
    0.001655 \; {\cal I} (R_{10})+ 0.0005 \; { \cal I}(R_{10} R_{8}^*) +
    0.0535 \; {\cal I} (R_{7})+ 0.02266 \; { \cal I} (R_{7} R_{8}^*) \nonumber\\
& &
+ 0.00496\; { \cal I}( R_{7} R_{9}^*)+ 0.00513 \; {\cal I}
(R_{8})+ 0.0261 \; {\cal I} (R_{8}
R_{9}^*)- 0.0118 \; {\cal I} (R_{9}) \nonumber\\
& & - 0.5426 \; {\cal R} (R_{10})+ 0.0281 \; {\cal R} (R_{7})+
0.0153 \; { \cal R}( R_{7} R_{10}^*)+ 0.06859 \;  { \cal R} (R_{7}
R_{8}^*) \nonumber\\ & & - 0.8554 \; { \cal R} (R_{7} R_{9}^*)-
0.00866 \; {\cal R} (R_{8}) + 0.00185 \; { \cal R}( R_{8} R_{10}^*)-
0.0981 \; { \cal R} (R_{8} R_{9}^*) \nonumber\\ & & + 2.7008 \;
{\cal R} (R_{9})- 0.10705 \; {\cal R}( R_{9} R_{10}^*)+ 10.7687 \;
|R_{10}|^2+ 0.2889 \; |R_7|^2 \nonumber\\ & & + 0.00381 \; |R_8|^2 +
1.4892 \; |R_9|^2\; \Big] \times 10^{-7} \; .
\end{eqnarray}
where $R_i \equiv C_i(\mu_0)/C_i^{\rm SM} (\mu_0)$. The SM prediction
is $BR(B\to X_s \ell^+ \ell^-) = (1.59 \pm 0.11)10^{-6}$ and we will
assign a theoretical error of 6.9\% to the central values calculated
in Eq.~(\ref{bsll}).

The impact that the $B\to X_s \gamma$ and $B\to X_s \ell\ell$
measurements have on the T2HDM parameter space is shown in
Figs.~\ref{fig:bsg} and \ref{fig:bsee}. In Fig.~\ref{fig:bsg}a we plot
the $B\to X_s \gamma$ branching ratio as a function of the charged
Higgs mass for various choices of $\tan\beta_H$ and $\xi$. The
$\tan\beta_H$ dependence of the charged Higgs contributions to $C_7$ is
not very strong as it follows from the proximity of the solid and
dashed curves. The $\xi$ dependence is, on the other hand, much
stronger; here we plot results for $\xi=(1,-1)$ (other choices of the
phase yield in between curves). This can be seen explicitly in
Fig.~\ref{fig:bsg}b, where we plot the allowed region at 68\% C.L. in
the $(\tan\beta_H,m_{H^\pm})$ plane for various choices of
$\xi$. Comparison of the plots in Figs.~\ref{fig:bsg} and
\ref{fig:bsee} shows that $B\to X_s \ell\ell$ does not provide
additional constraints on the parameter space.

\begin{table}
\begin{center}
\begin{displaymath}
\begin{tabular}{|l|l|}
\hline
\spp $G_F = 1.16639 10^{-5} \; \gev^{-2}$ &
     $\lambda = 0.2258 \pm 0.0014$~\cite{Bona:2006ah} \\
\spp $m_W = 80.426\;\gev$ &
     $ A = 0.818 \pm 0.012$~\cite{Bona:2006ah}  \\
\spp $m_K = 0.497648\; \gev$  &
     $\bar \rho = 0.197 \pm 0.031$~\cite{Bona:2006ah}  \\
\spp $m_c(m_c) = (1.224 \pm 0.017 \pm 0.054)\;\gev$ \cite{Hoang:2005zw}  &
     $\bar\eta = 0.351 \pm 0.020$~\cite{Bona:2006ah} \\
\spp $m_{t,{\rm pole}}= (171.4 \pm 2.1) \;\gev$ \cite{topmass} &
     $\alpha_s^{\overline{MS}}(m_Z)=0.1182 \pm 0.0027$ \cite{Bethke:2004uy}\\
\spp $\sin^2 \theta_W =0.2312$ &
     $m_Z=91.1876 \; \gev$ \\
\spp $m_b^{1S} = (4.68 \pm 0.03)\; \gev$ \cite{Bauer:2004ve}&
     \\ \hline
\end{tabular}
\end{displaymath}
\caption{Numerical inputs that we use in the phenomenological
  analysis. Unless explicitly specified, they are taken from the
  PDG~\cite{Yao:2006px}.\label{tab:inputs}}
\end{center}
\end{table}
%

%%%%%%%%%%%%%%%%%%%%%%%%%%%%%%%%%%%%%%%%%%%%%%%%%%%%%%%%%%%%%%%%%%%%%%%
%%%%%%%%%%%%%%%%%%%%%%% FIGURE: DMK  %%%%%%%%%%%%%%%%%%%%%%%%%%%%%%%%%%
%%%%%%%%%%%%%%%%%%%%%%%%%%%%%%%%%%%%%%%%%%%%%%%%%%%%%%%%%%%%%%%%%%%%%%%
\begin{figure}
\begin{center}
\includegraphics[width=0.48 \linewidth]{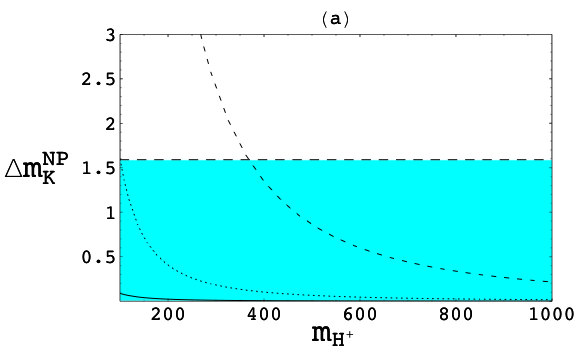}
\raisebox{0.1cm}{\includegraphics[width=0.48 \linewidth]{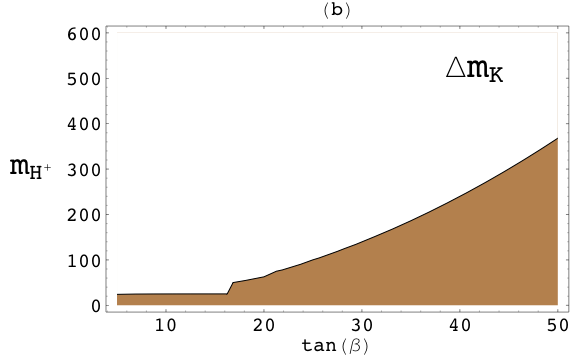}}
\end{center}
\vskip -0.7cm
\caption{
{\bf Plot a}. $m_{H^\pm}$ dependence of the T2HDM contributions to
$\Delta m_K$ in units of $10^{-3} \; {\rm ps}^{-1}$. Solid, dotted and
dashed lines correspond to $\tan\beta_H=10$, $25$ and $50$,
respectively. There is no appreciable dependence on $\xi$ and
$\xi^\prime$. The horizontal dashed line corresponds to $\Delta
m_K^{\rm NP} < 0.3 \; \Delta m_K^{\rm exp}$.
{\bf Plot b}. Portion of the $(\tan\beta_H,m_{H^\pm})$ plane excluded by
the $\Delta m_K^{\rm NP} < 0.3 \; \Delta m_K^{\rm exp}$ constraint.
}
\label{fig:dmk}
\end{figure}
\subsection{Neutral mesons mixing}
\label{sec:neutralmesonmixing}
The off-diagonal element of the neutral K-mesons mass matrix is
$ M_{12}^* = \langle \overline K^0 \left| {\cal H}_{\rm eff} \right| K^0 \rangle/ (2 m_K)$,
where the effective Hamiltonian is
\bea
{\cal H}_{\rm eff} & = & \frac{G_F^2 m_W^2}{16 \pi^2}
\left( V_{ts}^* V_{td}^{} \right)^2 \sum_a C_a (\mu) Q_a \; ,
\eea
with
\begin{eqnarray}\label{OPS}
&&Q^{\rm VLL}=\left(\overline s_L\gamma_\mu d_L\right)
\left(\overline s_L \gamma^\mu d_L\right)\nonumber\\
&&Q_1^{\rm LR}=\left(\overline s_L\gamma_\mu d_L\right)
\left(\overline s_R\gamma_\mu d_R\right)\nonumber\\
&&Q_2^{\rm LR}=\left(\overline s_R d_L\right)
\left(\overline s_L d_R\right) \label{eqn:ops_dF2}\\
&&Q_1^{\rm SLL}=\left(\overline s_R d_L\right)
\left(\overline s_R d_L\right) \nonumber\\
&&Q_2^{\rm SLL}=\left(\overline s_R \sigma_{\mu\nu} d_L\right)
\left(\overline s_R \sigma^{\mu\nu} d_L\right) \; . \nonumber
\end{eqnarray}
The additional operators $Q^{\rm VRR}$, $Q_1^{\rm SRR}$ and
$Q_2^{\rm SRR}$ are obtained from $Q^{\rm VLL}$, $Q_1^{\rm SLL}$ and
$Q_2^{\rm SLL}$ by replacing $L$ with $R$. The effective
Hamiltonians that describe $B$ and $B_s$ mixing are obtained via the
replacements $(s,d) \to (b,d)$ and $(s,d) \to (b,s)$, respectively.
The $D$ mixing Hamiltonian requires $(s,d)\to (c,u)$ and $V_{ts}^*
V_{td}^{} \to V_{cb}^* V_{ub}^{}$).

In the SM only the coefficient $C^{\rm VLL}$ receives sizable
contributions (in the D meson sector the GIM cancelation is more
effective due to the smallness of the b quark with respect to the
top one).

\begin{table}
\begin{center}
%\begin{displaymath}
\begin{tabular}{|l|l|l|}
\hline
\spp $r=0.985$~\cite{Buras:2001mb}  &
     $ \eta_1 = 1.32 \; \left(\frac{1.3}{m_c(m_c)}\right)^{1.1} \pm 0.32$~\cite{Battaglia:2003in}  &
     $\hat B_K=0.79 \pm 0.04 \pm 0.08$~\cite{Bona:2006ah}\\
\spp $f_K = 0.159\; \gev$~\cite{Bona:2006ah} &
     $\eta_2 = 0.57 \pm 0.01$~\cite{Buras:1998ra} &
     $\eta_3 = 0.47 \pm 0.05$~\cite{Buras:1998ra} \\
\spp $P_{1,K}^{LR} = -36.1 $~\cite{Buras:2001ra} &
     $P_{2,K}^{LR} = 59.3 $~\cite{Buras:2001ra} &
     $P_{1,K}^{SLL} = -18.1 $~\cite{Buras:2001ra} \\
\spp $P_{2,K}^{SLL} = -32.2 $~\cite{Buras:2001ra} &
     $\Delta m_K^{\rm exp} = (5.301 \; 10^{-3}) \; {\rm ps}^{-1}$ &
     $\varepsilon_K^{\rm exp} = (2.280 \pm 0.013 ) \; 10^{-3}$  \\
\hline
\end{tabular}
%\end{displaymath}
\caption{Inputs that we use in the phenomenological analysis of
  $K-\bar K$ mixing.\label{tab:latticeKmixing}}
\end{center}
\end{table}
\begin{table}
\begin{center}
%\begin{displaymath}
\begin{tabular}{|l|l|l|}
\hline
\spp $f_{B_s} \sqrt{\hat B_s} = (0.281 \pm 0.021) \; \gev$~\cite{Dalgic:2006gp} &
     $m_{B_s} = 5.36675 \; \gev$  &
     $m_{B_d} = 5.2794 \; \gev$ \\
\spp $f_{B_s}/f_{B_d} = 1.20 \pm 0.03$~\cite{Mackenzie:2006un} &
     $P_{1,B_d}^{LR} = -0.89$ &
     $P_{1,B_s}^{LR} = -0.98$ \\
\spp $m_s^{\overline{MS}} (2\; \gev) = (0.076 \pm 0.08) \; \gev$~\cite{Aubin:2004fs} &
     $P_{2,B_d}^{LR} = 1.13$&
     $P_{2,B_s}^{LR} = 1.24$ \\
\spp $f_{B_d} = (0.216 \pm 0.022) \; \gev$~\cite{Mackenzie:2006un} &
     $P_{1,B_d}^{SLL} = -0.46$ &
     $P_{1,B_s}^{SLL} = -0.51$ \\
\spp $\xi_s = f_{B_s}/f_{B_d} \sqrt{\hat B_s/\hat B_d}
            = 1.210^{+0.047}_{-0.035}$~\cite{Mackenzie:2006un} &
     $P_{2,B_d}^{SLL} = -0.90$ &
     $P_{2,B_s}^{SLL} = -0.98$ \\
\spp $\Delta m_{B_d}^{\rm exp} = (0.507 \pm 0.005) {\rm ps}^{-1}$ &
     $\eta_B=0.55$~\cite{Buras:2001mb} &
     $a_{\psi K_s}^{\rm exp} = 0.675 \pm 0.026$ \\
\spp $\Delta m_{B_s}^{\rm exp} = (17.77 \pm 0.10 \pm 0.07) {\rm ps}^{-1}$ &
      &
      \\
\hline
\spp $f_{D} = 0.165 \; \gev$~\cite{Petrov:1997ch} &
     $m_{D} = 1.8645 \; \gev$  &
     $B_D = 0.78 \pm 0.01$~\cite{Gupta:1996yt,Lin:2006vc}\\
\hline
\end{tabular}
%\end{displaymath}
\caption{Inputs that we use in the phenomenological analysis of
  $B_q-\bar B_q$ and $D-\bar D$ mixing.\label{tab:latticeBDmixing}}
\end{center}
\end{table}

In the T2HDM there are no tree-level flavor changing neutral Higgs
currents involving down quarks; hence the Wilson coefficients for
$K$, $B$ and $B_s$ mixing receive non standard contributions only
through charged Higgs box diagrams. The latter can be found, for
instance, in Eq.~(A.11) of Ref.~\cite{Buras:2002vd}~\footnote{We
defined the couplings $P^{H,G}_{LR,RL}$ in Eq.~(\ref{Lcharged}) in
complete analogy to Ref.~\cite{Buras:2002vd}}.

The situation is different for what concerns $D-\bar D$ mixing. In
fact, from Eqs.~(\ref{Lneutral}) and (\ref{sigma}), it follows that
the $\bar u_L c_R S^0$ ($S=h,\; H,\; A$) coupling is non-vanishing
(albeit quite small); therefore, it induces a tree level contribution
to the Wilson coefficient $C_1^{SLL}$. The charged Higgs box diagram
contributions are obtained from Eq.~(A.11) of Ref.~\cite{Buras:2002vd}
with the following replacements: $d \to u$, $P_A^B \to
(P_A^B)^\dagger$ (for $A=LR,RL$ and $B=G,H$), $V\to V^\dagger$, $LR
\leftrightarrow RL$ and $(ji)\to (21)$. Neutral Higgs box diagrams
involve the small $\bar u_L c_R S^0$ coupling and are suppressed with
respect to the tree level contributions.

%%%%%%%%%%%%%%%%%%%%%%%%%%%%%%%%%%%%%%%%%%%%%%%%%%%%%%%%%%%%%%%%%%%%%%%
%%%%%%%%%%%%%%%%%%%%%%% FIGURE: EpsilonK  %%%%%%%%%%%%%%%%%%%%%%%%%%%%%
%%%%%%%%%%%%%%%%%%%%%%%%%%%%%%%%%%%%%%%%%%%%%%%%%%%%%%%%%%%%%%%%%%%%%%%
\begin{figure}
\begin{center}
\includegraphics[width=0.48 \linewidth]{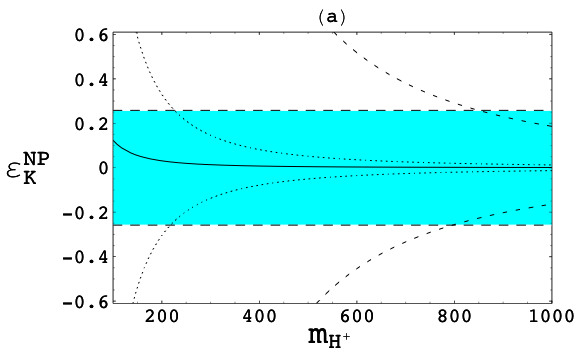}
\raisebox{0.1cm}{\includegraphics[width=0.46 \linewidth]{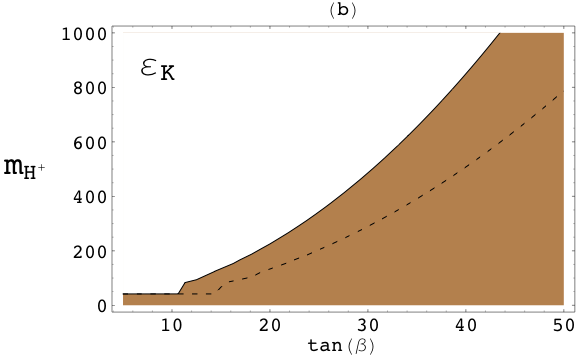}}
\end{center}
\vskip -0.7cm
\caption{
{\bf Plot a}. $m_{H^\pm}$ dependence of the T2HDM contributions to
$\varepsilon_K$ in units of $10^{-3}$ ($\varepsilon_K^{\rm NP} \equiv
\varepsilon_K^{\rm T2HDM} - \varepsilon_K^{\rm SM}$). Solid, dotted
and dashed lines correspond to $|\xi|=1$ and $\tan\beta_H=10$, $20$ and
$40$, respectively. Curves with $\varepsilon_K^{\rm NP}$ positive and
negative correspond to $\xi=(1,-1)$, respectively. There is no
appreciable dependence on $\xi^\prime$. The meaning of the blue region
is explained in the text.
{\bf Plot b}. Portion of the $(\tan\beta_H,m_{H^\pm})$ plane excluded by
$\varepsilon_K$. The shaded area corresponds to $\xi=1$. The dashed
line show how this region changes for $\xi=e^{i \pi/4}$. Other choices
of the phase yield in-between lines.
}
\label{fig:ek}
\end{figure}
\subsubsection{$K\overline K$ mixing}
\label{subsec:Kmixing}
The $K-\overline K$ mass difference and the measure of indirect
CP violation in the $K$ system are given by (see for instance Ref.~\cite{Buras:1998ra})
\bea
\Delta m_K & = & 2 \; {\rm Re} ( M_{12}^K ) \;, \\
\varepsilon_K & \equiv & \frac{A(K_L \to (\pi\pi)_{I=0})}{A(K_S \to (\pi\pi)_{I=0})}
                =        \frac{{\rm exp} (i\pi/4)}{\sqrt{2} \Delta m_K} \; {\rm Im}  ( M_{12}^K ) \;.
\eea
The expression for $M_{12}^K$ in presence of arbitrary new physics
contributions is~\cite{Buras:1998ra,Buras:2001ra,Buras:2001mb}:
\bea
(M_{12}^K)^* & = & \frac{G_F^2}{12\pi^2} f_K^2 \hat{B}_K m_K m_W^2
               \left[
                     \lambda_c^{*2} \eta_1 S_0(x_c) + \lambda_t^{*2} \eta_2 F_{tt}^K
                     + 2 \lambda_c^{*} \lambda_t^{*} \eta_3 S_0(x_c,x_t)
               \right] \;, \\
F_{tt}^K & = & \left[ S_0(x_t) + \frac{1}{4r} C_{\rm new,K}^{VLL} \right]
               + \frac{1}{4r} C_{1,K}^{VRR} + \bar{P}_{1,K}^{LR} C_{1,K}^{LR} + \bar{P}_{1,K}^{LR} C_{1,K}^{LR} \nn \\
         &   & + \bar{P}_{1,K}^{SLL} \left[ C_{1,K}^{SLL} +  C_{1,K}^{SRR} \right]
               + \bar{P}_{2,K}^{SLL} \left[ C_{2,K}^{SLL} +  C_{2,K}^{SRR} \right] \;,
\label{FttK}
\eea
where $\lambda_i = V_{is}^* V_{id}^{}$, $x_t=M_t^2/m_W^2$, $x_c =
M_c^2/m_W^2$, the functions $S_0$ are given for instance in
Ref.~\cite{Buras:1998ra}, $\eta_i$ and $r$ are the QCD correction to
$S_0(x_t)$ in the SM, $f_K$ is the kaon decay constant, $\hat B_K$ and
$\bar{P}_i^A \equiv P_i^A/(4 \eta_2 \hat B_K)$ are lattice QCD
determinations of the matrix elements of the operators in
Eq.~(\ref{OPS})~\cite{Buras:2001ra,Buras:2001mb}.  The numerical
inputs the we use are summarized in
Tables~\ref{tab:inputs}-\ref{tab:latticeKmixing}.

The $K\overline K$ mass difference receives additional long distance
contributions; in the numerical analysis we assume that such
non-perturbative effects do not contribute to more than $30\%$ of the
observed mass splitting (i.e. $(\Delta m_K)_{NP} < 0.00159 \; {\rm
ps}^{-1}$). See Ref.~\cite{Bijnens:1990mz} for an estimation of these
long distance effects in the large $N_c$ limit. An approximate
expression for $\Delta m_K$ is the following:
\bea
\Delta m_K \simeq \frac{G_F^2}{6 \pi^2}f_K^2 \hat B_K m_K {\rm Re} (\lambda_c^{*2}) \left(
           \eta_1 \left[m_c (m_c)\right]^2
         + \frac{\eta_2}{r} \frac{\left[m_c(m_t)\right]^4
                 \tan^4 \beta_H}{4 \; m^2_{H^\pm}} \right) \; .
\eea
Imposing the $(\Delta m_K)_{NP} < 0.00159\; {\rm ps}^{-1}$ constraint,
we obtain: $m_{H^\pm} > 89 \; ( \tan\beta_H / 25 )^2 \; \gev$.

The exact numerical impact of the upper limit on $(\Delta m_K)_{NP}$
can be seen in Fig.~\ref{fig:dmk}. Comparison with Fig.~\ref{fig:bsg}b
shows that for $\xi>0$, this constraint is complementary to $B\to X_s
\gamma$.

The impact of the $\varepsilon_K$ measurement is shown in
Fig.~\ref{fig:ek}. Here we require $\varepsilon_K^{\rm T2HDM}$ to lie
in the $\varepsilon_K$ range extracted from the standard unitarity
triangle analysis. A more correct approach is to fit the unitarity
triangle in the T2HDM and check whether each given point in the
parameter space gives an acceptable chi-square. This analysis is
presented in Sec.~\ref{sec:chisquare}.

We find that the inclusion of the $\varepsilon_K$ constraint has a
very strong impact. Note that, in this case, the effect is
proportional to $\xi$; hence, $B\to X_s \gamma$ is still required in
the $\xi \sim 0$ limit.

\subsubsection{$B_q\overline B_q$ mixing}
\label{subsec:Bmixing}

%%%%%%%%%%%%%%%%%%%%%%%%%%%%%%%%%%%%%%%%%%%%%%%%%%%%%%%%%%%%%%%%%%%%%%%
%%%%%%%%%%%%%%%%%%%%%%% FIGURE: DeltaMBs %%%%%%%%%%%%%%%%%%%%%%%%%%%%%%
%%%%%%%%%%%%%%%%%%%%%%%%%%%%%%%%%%%%%%%%%%%%%%%%%%%%%%%%%%%%%%%%%%%%%%%
\begin{figure}
\begin{center}
\includegraphics[width=0.48 \linewidth]{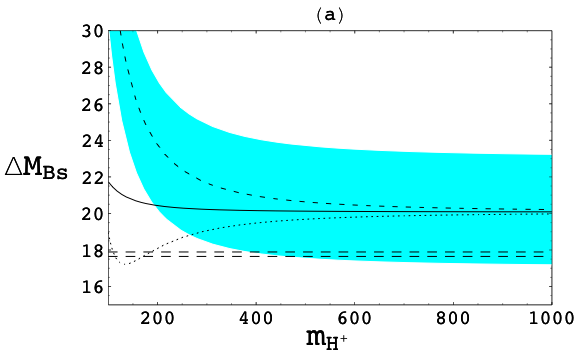}
\raisebox{0.1cm}{\includegraphics[width=0.46 \linewidth]{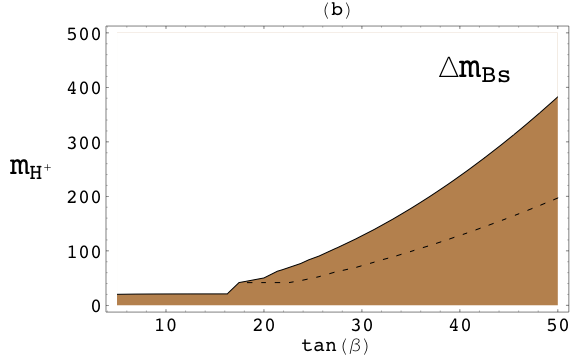}}
\includegraphics[width=0.48 \linewidth]{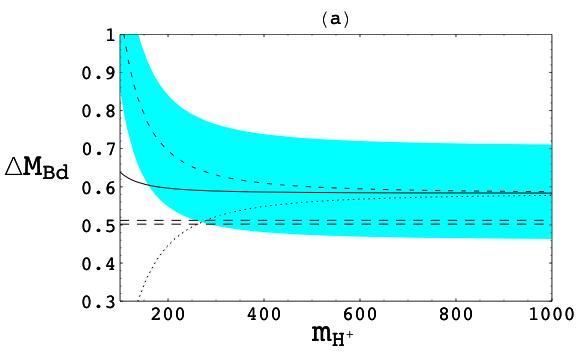}
\raisebox{0.1cm}{\includegraphics[width=0.46 \linewidth]{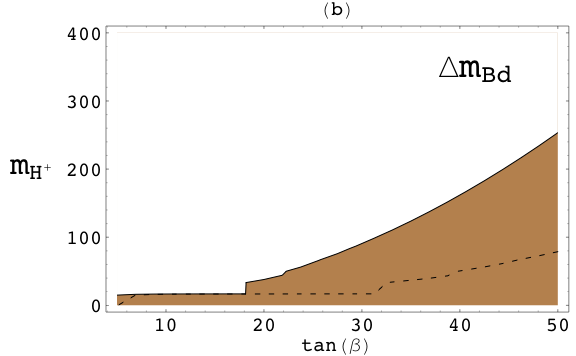}}
\end{center}
\vskip -0.7cm
\caption{
{\bf Plots a}. $m_{H^\pm}$ dependence of the T2HDM contributions to
$\Delta m_{B_{(s,d)}}$ in ${\rm ps}^{-1}$. Solid, dotted and dashed
lines correspond to $|\xi|=1$ and $(\tan\beta_H,\varphi_\xi)=$ $(30,0)$,
$(50,0)$, $(50,\pi/2)$, respectively. There is no appreciable
dependence on $\xi^\prime$. The horizontal dashed lines are the
experimental measurement. The blue band shows the theoretical
uncertainties for the dashed line, similar bands can be drawn for the
other curves.
{\bf Plots b}. Portions of the $(\tan\beta_H,m_{H^\pm})$ plane excluded
by $\Delta m_{B_{(s,d)}}$. The shaded area corresponds to $\xi=1$. The
dashed line show how this region changes for $\xi=e^{i \pi/2}$. Other
choices of the phase yield in-between lines.
}
\label{fig:dmb}
\end{figure}

The $B_q-\bar B_q$ mass difference is given by~\cite{Buras:2002vd}
\bea
\Delta m_{B_q} & = & \frac{G_F^2 m_W^2}{6\pi^2} m_{B_q} \eta_B f_{B_q}^2 \hat B_{B_q}
                     \left| V_{tb}^{} V_{tq}^*\right|^2 \left|F_{tt}^{B_q}\right|
\eea
where $F_{tt}^{B_q}$ is given by Eq.~(\ref{FttK}) with the replacement
$K\rightarrow B_q$. We recalculated the quantities $\bar{P}_i^A \equiv
P_i^A/(4 \eta_B \hat B_{B_q})$ using the formulae presented in
Ref.~\cite{Buras:2001ra} and the lattice results of
Ref.~\cite{Becirevic:2001xt,Becirevic:2001yv}. The numerical inputs
that we use are collected in Table~\ref{tab:latticeBDmixing}.

The SM prediction for $\Delta m_{B_s}$ does not depend on the
extraction of the CKM parameters $\rho$ and $\eta$; using the inputs
summarized in Table~\ref{tab:latticeBDmixing}, we obtain $\Delta
m_{B_s}^{SM} = (20.5 \pm 3.1) \; {\rm ps}^{-1}$. Note that, in the SM,
it is possible to use the measurement of $\Delta m_{B_d}$ to obtain a
second determination of $f_{B_d} \sqrt{B_d}$ and of $f_{B_s}
\sqrt{B_s}$ (via $\xi_s$), thus reducing the error on the prediction
for $\Delta M_{B_s}$.

The situation for $\Delta m_{B_d}$ is different. From inspection of
the standard fits of the unitarity triangle, it is clear that is
always possible to choose $\rho$ and $\eta$ such that the SM
prediction agrees perfectly with the experimental central value. For
this reason, in the numerical, analysis we just require the new
physics contributions to $\Delta m_{B_d}$ to be compatible with the
experimental determination up to an uncertainty given by the lattice
errors on $f_{B_d} \sqrt{B_d}$. A more correct analysis requires a
simultaneous fit of the new physics contributions to $\varepsilon_K$,
$\Delta m_{B_q}$, $a_{\psi K_s}$ and $|V_{ub}/V_{cb}|$. See
Ref.~\cite{Ball:2006xx} for a general discussion of New Physics
effects on $B_s$ mixing.

From the plots in Fig.~\ref{fig:dmb} we see that $B_q - \bar B_q$
mixing data constraints are still much weaker than the corresponding
constraint on $\varepsilon_K$.

%%%%%%%%%%%%%%%%%%%%%%%%%%%%%%%%%%%%%%%%%%%%%%%%%%%%%%%%%%%%%%%%%%%%%%%
\subsubsection{$D\overline D$ mixing}
\label{subsec:Dmixing}
%%%%%%%%%%%%%%%%%%%%%%%%%%%%%%%%%%%%%%%%%%%%%%%%%%%%%%%%%%%%%%%%%%%%%%%
%
The SM prediction for $\delta m_D$ range between $10^{-6} \; {\rm
ps}^{-1}$ and $10^{-2}\; {\rm ps}^{-1}$ and is completely dominated by
long distance effects; in fact, the short-distance SM prediction has
been calculated and reads~\cite{Golowich:2005pt,Golowich:2007ka} $x_D
\simeq 1.5 \times 10^{-6} \; {\rm ps}^{-1}$. The present experimental
information on $D \bar D$ mixing
parameters~\cite{Aubert:2007wf,Abe:2007dt}, yields the following model
independent determination of the $D\bar D$ mass
difference~\cite{Ciuchini:2007cw}: $\Delta m_D = (14.5 \pm 5.6)
10^{-3} {\rm ps}^{-1}$. In the T2HDM very large effects are possible
(of order 1\%~\cite{Wu:1999nc}), and there is the possibility that the
actual $D-\bar D$ mass difference is entirely controlled by new
physics short distance effects. In the numerics we require the new
physics contribution to the $\Delta m_D$ not to exceed the
measurement.

%%%%%%%%%%%%%%%%%%%%%%%%%%%%%%%%%%%%%%%%%%%%%%%%%%%%%%%%%%%%%%%%%%%%%%%
%%%%%%%%%%%%%%%%%%%%%%% FIGURE: xsd %%%%%%%%%%%%%%%%%%%%%%%%%%%%%%%%%%%
%%%%%%%%%%%%%%%%%%%%%%%%%%%%%%%%%%%%%%%%%%%%%%%%%%%%%%%%%%%%%%%%%%%%%%%
\begin{figure}
\begin{center}
\includegraphics[width=0.48 \linewidth]{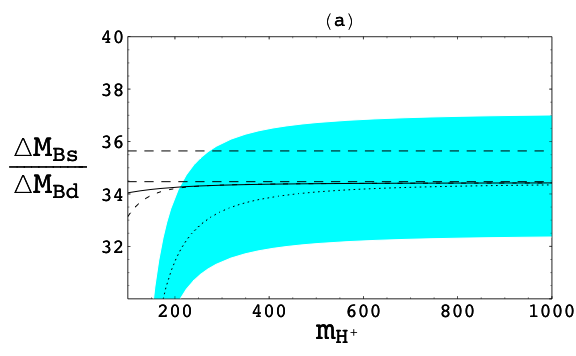}
\raisebox{0.1cm}{\includegraphics[width=0.46 \linewidth]{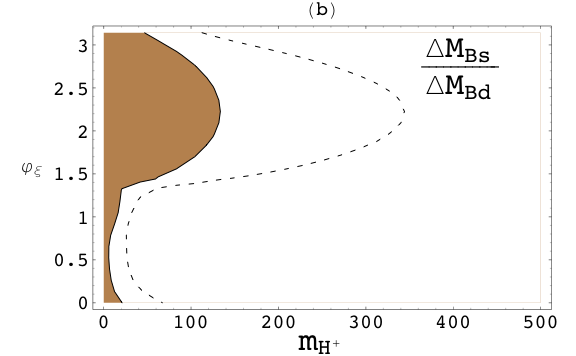}}
\end{center}
\vskip -0.7cm
\caption{
{\bf Plot a}. $m_{H^\pm}$ dependence of the T2HDM contributions to
$\Delta m_{B_{(s)}}/\Delta m_{B_{(d)}}$. See the caption in
Fig.~\ref{fig:dmb}.
{\bf Plot b}. Excluded region in the $(\varphi_\xi,m_{H^\pm})$
plane. The solid and dashed contours correspond to $\tan\beta_H=$30 and
50, respectively.}
\label{fig:xsd}
\end{figure}
%%%%%%%%%%%%%%%%%%%%%%%%%%%%%%%%%%%%%%%%%%%%%%%%%%%%%%%%%%%%%%%%%%%%%%%
%%%%%%%%%%%%%%%%%%%%%%% FIGURE: Delta MD %%%%%%%%%%%%%%%%%%%%%%%%%%%%%%
%%%%%%%%%%%%%%%%%%%%%%%%%%%%%%%%%%%%%%%%%%%%%%%%%%%%%%%%%%%%%%%%%%%%%%%
\begin{figure}
\begin{center}
\includegraphics[width=0.48 \linewidth]{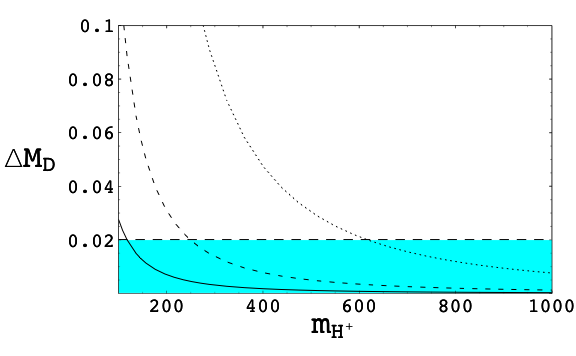}
\raisebox{0.1cm}{\includegraphics[width=0.46 \linewidth]{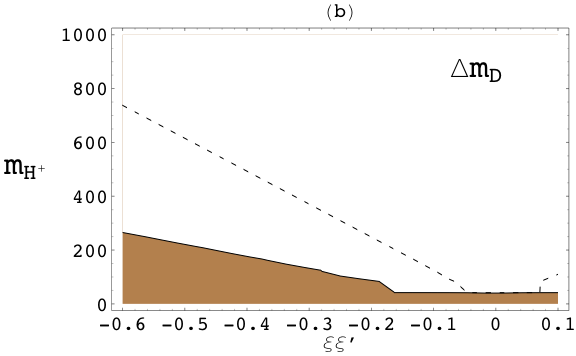}}
\end{center}
\vskip -0.7cm
\caption{
{\bf Plot a}. $m_{H^\pm}$ dependence of the T2HDM contributions to
$\Delta m_D$. Solid, dashed and dotted lines correspond to
$|\xi\xi^\prime|=0.1$, 0.2 and 0.5, respectively. We fix
$\tan\beta_H=50$. The horizontal dashed line is the experimental upper limit.
{\bf Plot b}. Portion of the $(\xi \xi^\prime ,m_{H^\pm})$ plane
excluded by $\Delta m_D$. The shaded area corresponds to
$\tan\beta_H=30$.  The dashed line to $\tan\beta_H=50$.
}
\label{fig:dmd}
\end{figure}

The $D-\bar D$ mass difference is given by
\bea
\Delta m_D & = &
\frac{G_F^2 m_W^2}{16\pi^2 m_D} \left| V_{ub}^{} V_{cq}^*\right|^2 \Bigg[
( C_{1,D}^{VLL} + C_{1,D}^{VRR} ) \; \langle Q^{VLL}   \rangle +
( C_{1,D}^{SLL} + C_{1,D}^{SRR} ) \; \langle Q_1^{SLL} \rangle + \nonumber \\
&& ( C_{2,D}^{SLL} + C_{2,D}^{SRR} ) \; \langle Q_2^{SLL} \rangle +
  C_{1,D}^{LR} \; \langle Q_1^{LR} \rangle +
  C_{2,D}^{LR} \; \langle Q_2^{LR} \rangle
\Bigg]
\eea
where the matrix elements are
\bea
\langle D| Q^{VLL} | \bar D \rangle & = & \frac{2}{3} \; m_D^2 f_D^2 \hat B^{VLL} \\
\langle D| Q_1^{SLL} | \bar D \rangle & = & - \frac{5}{12} R \; m_D^2 f_D^2 \hat B_1^{SLL} \\
\langle D| Q_2^{SLL} | \bar D \rangle & = & - R \; m_D^2 f_D^2 \hat B_2^{SLL} \\
\langle D| Q_1^{LR} | \bar D \rangle & = & - \frac{1}{3} R \; m_D^2 f_D^2 \hat B_1^{LR} \\
\langle D| Q_2^{LR} | \bar D \rangle & = & \frac{1}{2} R \; m_D^2 f_D^2 \hat B_2^{LR}
\eea
and $R = (m_D/(m_c + m_u))^2$. In the numerical analysis we use $\hat
B^{VLL} = \hat B_D = 0.82\pm 0.01$~\cite{Golowich:2005pt} (this value
of the {\it hat} parameter $\hat B_D$ has been obtained from the
lattice determination of $B_D (2\gev)$~\cite{Gupta:1996yt,Lin:2006vc})
and set all the other $B$ parameters to 1.

An approximate expression for the $D-\bar D$ mass difference is given
by~\cite{Wu:1999nc}
\bea
\Delta m_D & = & \frac{G_F^2}{6 \pi^2} (\xi \xi^{\prime *})^2
\frac{m_c^4 \tan^4\beta_H}{4 m_{H^\pm}^2} \; .
\eea
The strong $\xi^\prime$ dependence implies that, once $B\to \tau \nu$
data are imposed, no large deviations can be observed on $\Delta m_D$
as can be seen from Fig~\ref{fig:dmd}.

%%%%%%%%%%%%%%%%%%%%%%%%%%%%%%%%%%%%%%%%%%%%%%%%%%%%%%%%%%%%%%%%%%%%%%%
%%%%%%%%%%%%%%%%%%%%%%% FIGURE: B->tau nu %%%%%%%%%%%%%%%%%%%%%%%%%%%%%
%%%%%%%%%%%%%%%%%%%%%%%%%%%%%%%%%%%%%%%%%%%%%%%%%%%%%%%%%%%%%%%%%%%%%%%
\begin{figure}
\begin{center}
\includegraphics[width=0.48 \linewidth]{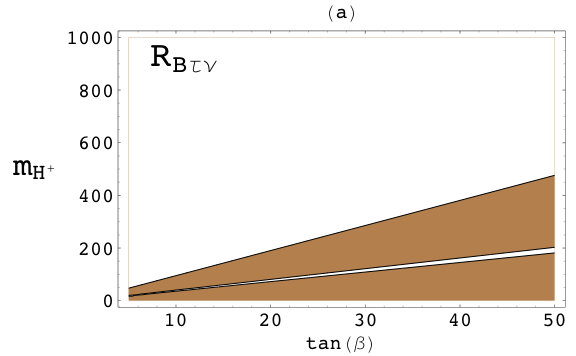}
\raisebox{0.1cm}{\includegraphics[width=0.46 \linewidth]{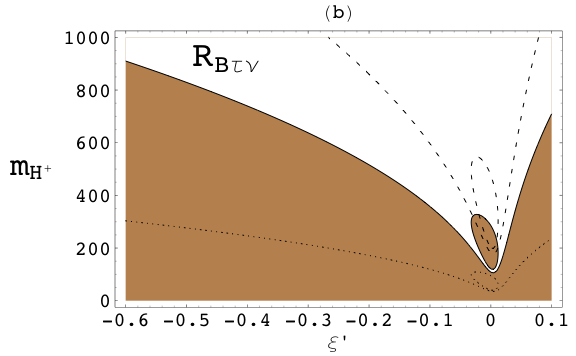}}
\end{center}
\vskip -0.7cm
\caption{
{\bf Plot a}. Portion of the $(\tan\beta_H,m_{H^\pm})$ plane allowed by
$R_{B\tau \nu}$ for $\xi^\prime = 0$.
{\bf Plot b}. Excluded region in the $(\xi^\prime,m_{H^\pm})$
plane. The dotted, solid and dashed contours correspond to
$\tan\beta_H=$10, 30 and 50, respectively.}
\label{fig:rbtaunu}
\end{figure}

\subsection{$B^+ \to \tau^+ \nu_\tau$}
\label{subsec:btaunu}
The branching ratio for the decay $B\to \tau \nu_\tau$ has been
recently measured by the Belle~\cite{Ikado:2006un} and
Babar~\cite{Aubert:2006fk} collaborations
\bea
{\cal B} (B\to \tau \nu_\tau) & = &
\left( 1.79^{+0.56}_{-0.49}({\rm stat})^{+0.39}_{-0.46}({\rm syst}) \right) \times 10^{-4}
\;\; [{\rm Belle}] \\
{\cal B} (B\to \tau \nu_\tau) & = &
\left( 0.88^{+0.68}_{-0.67}({\rm stat}) \pm 0.11 ({\rm syst}) \right) \times 10^{-4}
\;\; [{\rm Babar}] \; ,
\eea
yielding the following world average
\bea
{\cal B}_{WA} (B\to \tau \nu_\tau) = ( 1.31 \pm 0.48 ) \times 10^{-4} \; .
\eea

The SM expectation reads:
\bea
{\cal B}_{\rm SM} (B\to \tau \nu_\tau)
& = & \frac{G_F^2 m_B m_\tau^2}{8\pi} \left( 1- \frac{m_\tau^2}{m_B^2} \right)
      f_B^2 |V_{ub}^{}|^2 \tau_B  =  (1.53 \pm 0.38) \times 10^{-4} \; ,
\eea
where we used the PDG world average $|V_{ub}| = (4.31 \pm 0.3 )\times
10^{-3}$ from direct tree level measurements only. The above result
leads to
\bea
R_{B\tau\nu} & = &
\frac{{\cal B}_{WA} (B\to \tau \nu_\tau)}{{\cal B}_{\rm SM} (B\to \tau \nu_\tau)}
 = 0.86 \pm 0.38 \; .
\label{rbtaunu}
\eea
If we use the fitted value of the CKM angles ($|V_{ub}|=(3.68\pm0.14 )
\times 10^{-3}$, the prediction reads $R_{B\tau\nu} = 1.18\pm
0.50$. The discrepancy between this determination of $R_{B\tau \nu}$ and
Eq.~(\ref{rbtaunu}) is a manifestation of the conflict within the SM
between the present determinations of $V_{ub}$ and $\sin (2\beta)$.

In the T2HDM this process receives large tree level contributions via
charged Higgs exchange:
\bea
{\cal B} (B\to \tau \nu_\tau)
& = &
{\cal B}_{\rm SM} (B\to \tau \nu_\tau)
\left| 1 - \tan^2 \beta_H \frac{m_B^2}{m^2_{H^\pm}}
\left( 1 - \frac{(\Sigma^\dagger V)_{13} }{m_b V_{ub}} \right) \right|^2 \\
& \simeq &
{\cal B}_{\rm SM} (B\to \tau \nu_\tau)
\left| 1 - \tan^2 \beta_H \frac{m_B^2}{m^2_{H^\pm}}
\left( 1 - \xi^{\prime *} \frac{m_c  V_{tb}}{m_b V_{ub}} \right) \right|^2 \; .
\label{rbtaunu1}
\eea

The numerical impact of the constraint in Eq.~(\ref{rbtaunu}) is very
strong. In Fig.~\ref{fig:rbtaunu}b we show the impact of this
constraint onto the $(\xi^\prime,m_{H^\pm})$ plane for various values
of $\tan\beta_H$. Since the experimental to SM ratio in
Eq.~(\ref{rbtaunu}) is smaller than 1, scenarios with $\xi^\prime >0$
are disfavored (see Eq.~(\ref{rbtaunu1})).

\subsection{Time-dependent CP asymmetry in $B\to (J/\psi,\phi,\eta^\prime) \; K_s$}
\label{sec:apsiks}
The time dependent CP asymmetry in the decay $B\to f K_s$
($f=J/\psi,\phi,\eta^\prime$) is given by
\bea
a_{f K} & = & \frac{2 \; {\rm Im} \lambda_{f K}}{1 + |\lambda_{f K}|^2} \\
\lambda_{\psi K} & = & - \frac{(M_{12}^{B_d})^*}{|M_{12}^{B_d}|}
                       \frac{A(\bar B^0 \to f K_s)}{A(B^0 \to f K_s)}
                  =  - e^{-2i\beta} \frac{F_{tt}^{B_d}}{|F_{tt}^{B_d}|}
                       \frac{A(\bar B^0 \to f K_s)}{A(B^0 \to f K_s)} \; ,
\eea
where $F_{tt}^{B_d}$ is obtained from Eq.~(\ref{FttK}) with obvious
replacements.  The effective Hamiltonian that controls the amplitude
$A(\bar B^0 \to f K_s)$ in the T2HDM is~\cite{Huber:2005ig}:
\bea
{\cal H}_{eff} = -\frac{4 G_F}{\sqrt{2}} V^*_{ts} V^{}_{tb}
\left[ \sum_{i=1}^{6} C_i(\mu) O_i(\mu) + \sum_{i=3}^{6} C_{iQ}(\mu) O_{iQ} (\mu)
+ C_{R} (\mu) \; O_{R} (\mu) \right]\;,
\label{heff}
\eea
where
\bea \label{ope}
O_1   & = & (\bar{s}_L \gamma_{\mu} T^a c_L) (\bar{c}_L \gamma^{\mu} T^a b_L), \\
O_2   & = & (\bar{s}_L \gamma_{\mu}     c_L) (\bar{c}_L \gamma^{\mu}     b_L),\\
O_3   & = & (\bar{s}_L \gamma_{\mu}     b_L) \sum (\bar{q}\gamma^{\mu}     q),\\
O_4   & = & (\bar{s}_L \gamma_{\mu} T^a b_L) \sum (\bar{q}\gamma^{\mu} T^a q), \\
O_5   & = & (\bar{s}_L \gamma_{\mu_1}
                     \gamma_{\mu_2}
                     \gamma_{\mu_3}    b_L)\sum (\bar{q} \gamma^{\mu_1}
                                                         \gamma^{\mu_2}
                                                         \gamma^{\mu_3}     q),  \\
O_6   & = & (\bar{s}_L \gamma_{\mu_1}
                     \gamma_{\mu_2}
                     \gamma_{\mu_3} T^a b_L)\sum (\bar{q} \gamma^{\mu_1}
                                                            \gamma^{\mu_2}
                                                            \gamma^{\mu_3} T^a q),\\
O_{3Q} & = & (\bar{s}_L \gamma_{\mu}     b_L) \sum Q_q (\bar{q}\gamma^{\mu}     q),\\
O_{4Q} & = & (\bar{s}_L \gamma_{\mu} T^a b_L) \sum Q_q (\bar{q}\gamma^{\mu} T^a q),\\
O_{5Q} & = & (\bar{s}_L \gamma_{\mu_1}
                     \gamma_{\mu_2}
                     \gamma_{\mu_3}    b_L)\sum Q_q (\bar{q} \gamma^{\mu_1}
                                                               \gamma^{\mu_2}
                                                               \gamma^{\mu_3}     q),\\
O_{6Q} & = & (\bar{s}_L \gamma_{\mu_1}
                     \gamma_{\mu_2}
                     \gamma_{\mu_3} T^a b_L)\sum Q_q (\bar{q} \gamma^{\mu_1}
                                                                \gamma^{\mu_2}
                                                                \gamma^{\mu_3} T^a q),\\
O_R & = & (\bar c_R  b_L) \; (\bar s_L  c_R).
\eea
Tree level and one--loop charged Higgs diagrams contribute to the
following matching conditions (we adopt the notation of
Ref.~\cite{Huber:2005ig}):
\bea
C_R^{(00)} (\mu_0) & = &
                - \frac{\left(4 V_{cb}^{} V_{cs}^{*} G_F/\sqrt{2} \right)^{-1}}{m_{H^\pm}^2}
               (P^H_{RL})_{23} \; (P^H_{RL})^*_{22}
                    \simeq  \frac{V_{tb}}{V_{cb}} \frac{m_c^2}{m_{H^\pm}^2} \xi^* \tan^2 \beta_H \; ,\\
\delta C_4^{(10)} (\mu_0) & = & E_H (x_{th}) \frac{\kappa}{\tan^2\beta_H} \; , \\
\delta C_3^{(11)} (\mu_0) & = & -\frac{2}{9  s^2_W} C_H (x_{th}) \frac{\kappa}{\tan^2\beta_H} \; \\
\delta C_5^{(11)} (\mu_0) & = & -\frac{1}{4} \delta C_3^{(11)} (\mu_0)  \; , \\
\delta C_{3Q}^{(11)} (\mu_0) & = & \left[ D_H (x_{th}) +
                                   4 \; C_H (x_{th}) \left( 1+\frac{1}{3 s^2_W} \right)
                    \right] \frac{\kappa}{\tan^2\beta_H} \; , \\
\delta C_{5Q}^{(11)} (\mu_0) & = & \frac{3}{2} \delta C_3^{(11)} (\mu_0)  \; ,
\eea
where $\mu_0 \sim O(m_t)$, $x_{th}= m_t^2/m_{H^\pm}^2$ and
\bea
\kappa & = & \frac{ (P^H_{RL})_{33} \; (P^H_{RL})^*_{32} \; \tan^2\beta_H}{
                    V_{tb}^{} V_{ts}^{*} \; m_t^2 \; 4 G_F/\sqrt{2}}
       \simeq  1 - \xi^* \tan^2\beta_H \frac{V_{cs}^*}{V_{ts}^*} \left( m_c\over m_t \right)^2
         \simeq 1 - 0.3 \; \xi^* \left( \tan\beta_H\over 30 \right)^2
\eea
The functions $E_H$, $D_H$ and $C_H$ can be found in Appendix A of
Ref.~\cite{Gabrielli:1994ff}. Note that the results for the type-II
2HDM are recovered in the $\kappa \to 1 $ limit.

Direct calculation of the anomalous dimensions involving the operator
$O_R$ yields (in the notation of Ref.~\cite{Huber:2005ig}):
\bea
\gamma_{RR}^{(10)} & =& -16 \; ,\\
\gamma_{R4}^{(10)} & =& -\frac{2}{3} \; ,\\
\gamma_{Rj}^{(10)} & =& 0 \; {\rm (j\neq R,4)}\; .\\
\eea
The large anomalous dimension of $O_R$ implies a large impact of the
running from the high-scale $\mu_0 \sim O(m_t)$ to the low-scale
$\mu_b \sim O(m_b)$:
\bea
C_R (\mu_b) & = & \eta^{-1.04348} \; C_R (\mu_0) \; ,
\eea
where $\eta = \alpha_s (\mu_0) /\alpha_s (\mu_b) \simeq 0.53$.

Let us first consider $A (\bar B^0 \to J/\psi K_s)$. The impact of the
QCD and electroweak penguin coefficients is very small; hence the only
T2HDM effect comes via the new operator $O_R$. Adopting the naive
factorization framework, the amplitude is proportional to:
\bea
A (\bar B^0 \to J/\psi K_s) & \propto &
\frac{4}{9} C_1 (\mu_b) + \frac{1}{3} C_2(\mu_b) + 2 \; C_3(\mu_b)
+ 20 \; C_5(\mu_b) + \frac{4}{3} C_{3Q}(\mu_b) + \frac{40}{3} C_{5Q}(\mu_b) \nn\\
& & - \frac{1}{6} C_R (\mu_b)\; ,
\eea
where the $C_R$ contribution receives a factor of $-1/2$ and $1/3$
from Lorentz and color Fierzing, respectively. Note that the SM
contribution (see for instance Ref.~\cite{Cheng:2000kt}) has been
rewritten in the new operator basis Eq.~(\ref{heff}). Direct
calculation of this amplitude in the QCD--factorization
approach~\cite{Cheng:2000kt} shows that the naive estimate is a fairly
good approximation.

The analysis of the amplitudes $A (\bar B^0 \to (\phi,\eta^\prime)
K_s)$ is more complicated. In this case the QCD and electroweak
penguin coefficients do play a leading role; moreover the magnetic
penguin operator $O_8$ contributes as well. In the following we adopt
the QCD-factorization analysis presented in
Ref.~\cite{Gabrielli:2004yi}. After transforming the Wilson
coefficients of Ref.~\cite{Gabrielli:2004yi} into our basis, we obtain
the following expressions for the $\phi$ and $\eta^\prime$ amplitudes:
\bea
A (\bar B^0 \to \eta^\prime K_s) & \propto &
(-0.04874 - 0.04905 i)
- (0.00658 - 0.00058 i) \; \delta C_3^{(11)} (\mu_0) \nn\\
& &
- (0.00014 - 0.00002 i) \; \delta C_{3Q}^{(11)} (\mu_0)
+ (0.00711 - 0.00275 i)  \; \delta C_4^{(10)} (\mu_0) \nn\\
& &
+ 0.0007 \; C_7 (\mu_b)
- 0.089 \; C_8 (\mu_b)
+ (0.03567 - 0.01087 i) \; C_R (\mu_0) \; , \\
A (\bar B^0 \to \phi K_s) & \propto &
(0.03262  + 0.00791 i)
+ (0.00963  - 0.00050 i)  \; \delta C_3^{(11)} (\mu_0) \nn\\
& &
+ (0.00044  - 0.00002 i) \; \delta C_{3Q}^{(11)} (\mu_0)
- (0.00282  - 0.00013 i) \; \delta C_4^{(10)} (\mu_0) \nn\\
& &
 - 0.0004\; C_7 (\mu_b) + 0.047\; C_8 (\mu_b) - (0.01292 - 0.00092 i)\; C_R (\mu_0) \; .
\eea
Note that the impact of the QCD and electroweak penguin matching
conditions is suppressed by an order of magnitude with the respect to
the leading contribution; in fact, the low--scale penguin Wilson
coefficients are dominated by the tree-level coefficient $C_2$ via the
RGE running. Not surprisingly, the effect of the other tree-level
operator ($O_R$) on the running is also very large.

The experimental measurements of these three asymmetries
read~\cite{Barberio:2006bi}:
\bea
a_{\psi K} & = & 0.675 \pm 0.026 \; ,\\
a_{\eta^\prime K} & = & 0.61 \pm 0.07 \; ,\\
a_{\phi K} & = & 0.39 \pm 0.18 \; .\\
\eea

\begin{figure}
\begin{center}
\includegraphics[width=0.48 \linewidth]{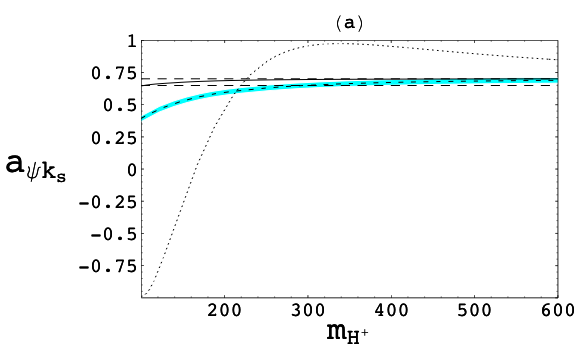}
\raisebox{0.1cm}{\includegraphics[width=0.46 \linewidth]{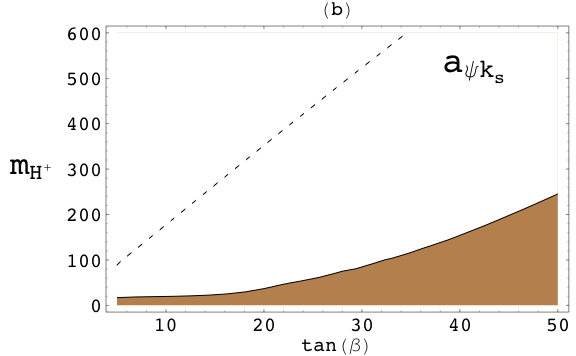}}
\includegraphics[width=0.48 \linewidth]{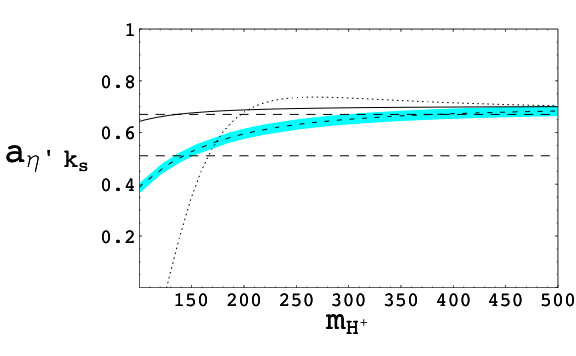}
\raisebox{0.1cm}{\includegraphics[width=0.46 \linewidth]{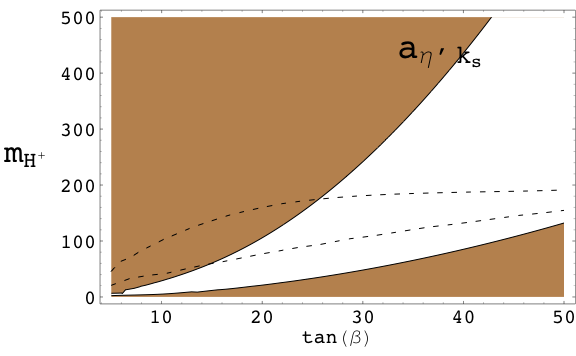}}
\includegraphics[width=0.48 \linewidth]{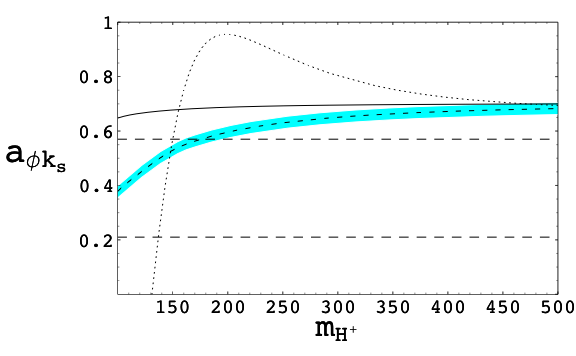}
\raisebox{0.1cm}{\includegraphics[width=0.46 \linewidth]{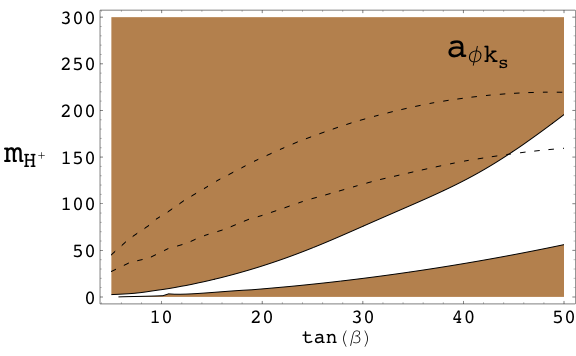}}
\end{center}
\vskip -0.7cm
\caption{
{\bf Plot a}. $m_{H^\pm}$ dependence of the T2HDM contributions to
$a_{(\psi,\eta^\prime,\phi) K}$. Solid, dotted and dashed lines
correspond to $|\xi|=1$ and $(\tan\beta_H,\varphi_\xi)=$ $(30,0)$,
$(50,0)$, $(50,\pi/2)$, respectively. There is no appreciable
dependence on $\xi^\prime$. The horizontal dashed lines are the
experimental measurement. The blue band shows the theoretical
uncertainties.
{\bf Plot b}. Portion of the $(\tan\beta_H,m_{H^\pm})$ plane excluded by
$a_{(\psi,\eta^\prime,\phi) K}$. The shaded area corresponds to
$\xi=1$. The dashed line show how this region changes for $\xi=e^{i
\pi/2}$ (in the first plot, the region excluded is below the dashed
line; in the second and third plots, it is above the uppermost dashed
line and below the lowermost one). Other choices of the phase yield
in-between contours.
}
\label{fig:apsiks}
\end{figure}

In Fig.~\ref{fig:apsiks}a, we show the size of T2HDM contributions
to the CP asymmetries in $B\to (J/\psi,\eta^\prime,\phi) K_S$ for
some choice of input parameters. In Fig.~\ref{fig:apsiks}b, we show
the portion of the ($\tan\beta_H,m_{H^\pm}$) parameter space that is
allowed by the present measurements of these asymmetries. From the
inspection of the figures we see that at the 1$\sigma$ level it is
possible to reconcile the $B\to \psi$ and $B\to \eta^\prime$
asymmetries in a quite wide region of the parameter space. The $B\to
\phi$ asymmetry, on the other hand, requires a too light charged
Higgs.

%%%%%%%%%%%%%%%%%%%%%%%%%%%%%%%%%%%%%%%%%%%%%%%%%%%%%%%%%%%%%%%%%%%%%%%
%%%%%%%%%%%%%%%%%%%%%%% FIGURE: apsiks %%%%%%%%%%%%%%%%%%%%%%%%%%%%%%%%
%%%%%%%%%%%%%%%%%%%%%%%%%%%%%%%%%%%%%%%%%%%%%%%%%%%%%%%%%%%%%%%%%%%%%%%

%%%%%%%%%%%%%%%%%%%%%%%%%%%%%%%%%%%%%%%%%%%%%%%%%%%%%%%%%%%%%%%%%%%%%%%
%%%%%%%%%%%%%%%%%%%%%%% FIGURE: Delta GAMMA %%%%%%%%%%%%%%%%%%%%%%%%%%%
%%%%%%%%%%%%%%%%%%%%%%%%%%%%%%%%%%%%%%%%%%%%%%%%%%%%%%%%%%%%%%%%%%%%%%%
\begin{figure}
\begin{center}
\includegraphics[width=0.48 \linewidth]{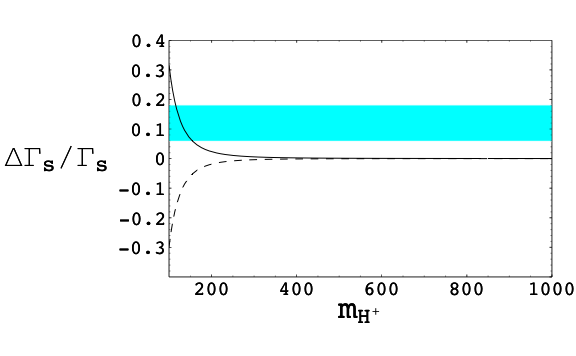}
\raisebox{0.1cm}{\includegraphics[width=0.46 \linewidth]{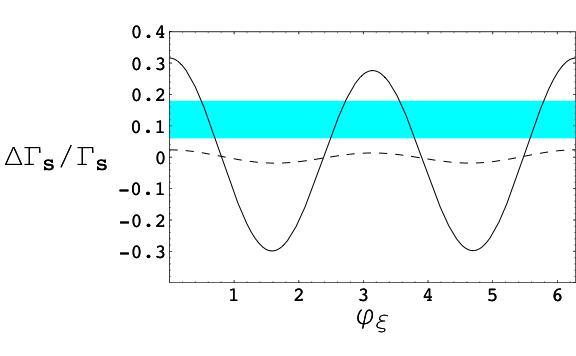}}
\end{center}
\vskip -0.7cm
\caption{
{\bf Plot a}. $m_{H^\pm}$ dependence of the T2HDM contributions to
$\Delta \Gamma_s/\Gamma_s$. We take $|\xi|=1$ and
$\tan\beta_H=50$. Solid and dashed lines correspond to $\varphi_\xi = 0$
and $\pi/2$, respectively. The blue band is the experimental 68\%
C.L. allowed region.
{\bf Plot b}. $\varphi_\xi$ dependence of the T2HDM contributions to
$\Delta \Gamma_s/\Gamma_s$. We take $|\xi|=1$ and
$\tan\beta_H=50$. Solid and dashed lines correspond to $m_{H^\pm}= 100$
and 200, respectively.  }
\label{fig:deltagamma}
\end{figure}

\subsection{$\Delta \Gamma_s / \Gamma_s$}
\label{subsec:deltagamma}
The $B_s$-$\bar B_s$ width difference is given by
\bea
\Delta \Gamma_s
&=&
- 2 \; \Gamma_{12}^s \cos (\beta_s + \theta_s)
 =  - 2 \Big([\Gamma_{12}^s]_{SM} + \delta \Gamma_{12}^s \Big) \cos (\beta_s + \theta_s)
 \\
\Gamma_{12}^s &=& \frac{1}{2 m_{B_s}} \langle \bar B_s |
{\rm Im} \left\{ i \int d^4 x \;  T \; {\cal H}_{\rm eff} (x) {\cal H}_{\rm eff} (0) \right\}
| B_s \rangle
\label{gamma12s}
\eea
where, in the Standard Model, ${\cal H}_{\rm eff}$ is the effective
Hamiltonian that mediates the bottom quark decay and is dominated by
tree level contributions. T2HDM contributions affect both $\theta_s$
and $\Gamma_{12}^s$: the former has been already discussed in
Sec.~\ref{subsec:Bmixing}; the latter are induced by a new $\Delta B =
1$ operator. The structure of the charged Higgs couplings in
Eqs.~(\ref{phlr}) and (\ref{phrl}), implies that the operator $O_{R}$
receives the largest contributions.  In the numerics we will consider
its effects together with the interference with the dominant Standard
Model operator $O_2$ (see Eq.~(\ref{heff}) for the definition of the
operators).

Direct calculation of the T-product in Eq.~(\ref{gamma12s}) yields the
following leading order expression for the new physics contribution to
$\Gamma_{12}^s$:
\bea
\delta \Delta\Gamma_{12}^s & = &
- \frac{G_F^2 m_{b,pole}^2}{12 \pi (2 m_{B_s})} \left( V_{cb}^* V_{cs}^{} \right)^2 \sqrt{1-4 z_c}
\Bigg[ \left(
\frac{1-4 z_c}{2} \frac{C_{RR}^2}{4} + \frac{C_2 C_{RR}}{2} {M_{c,pole}^2}{M_{b,pole}^2} \right)
\langle Q \rangle \nonumber \\
& & - (1 + 2 z_c) \frac{C_{RR}^2}{4} \langle Q_S \rangle
\Bigg]
\eea
where the operators $Q$ and $Q_S$ are
\bea
Q & = & (\bar b s)_{V-A} (\bar b s)_{V-A} \; ,\\
Q_S & = & (\bar b s)_{S-P} (\bar b s)_{S-P}
\eea
and their matrix elements between $\bar B_s$ and $B_s$ states are
\bea
\langle Q \rangle & = & f_{Bs}^2 m_{B_s}^2 \left( 1 + \frac{1}{N_c} \right) B \; , \\
\langle Q_S \rangle & = & -f_{Bs}^2 m_{B_s}^2 \frac{m_{B_s}^2}{(m_b + m_s)^2}
 \left( 2 - \frac{1}{N_c} \right) B_S \; ,
\eea
with $B = 0.87 \pm 0.06$ and $B_S = 0.84 \pm
0.05$~\cite{Ciuchini:2003ww}. After normalizing to the total $B_s$
width we obtain:
\bea
\frac{\Delta \Gamma_s}{\Gamma_s} & = & \tau_{B_s} \Delta \Gamma_s
= \left[ \frac{\Delta \Gamma_s}{\Gamma_s} \right]_{\rm SM}
  + \tau_{B_s} \; \delta \Delta \Gamma_s   \cos (\beta_s + \theta_s) \; .
\eea
The SM prediction~\cite{Lenz:2004nx,Lenz:2006hd} and the experimental
result~\cite{Yao:2006px,Abazov:2007tx} read:
\bea
\left[\frac{\Delta \Gamma_s}{\Gamma_s} \right]_{\rm SM}  & = & 0.147 \pm 0.060 \; , \\
\left[\frac{\Delta \Gamma_s}{\Gamma_s} \right]_{\rm exp} & = & 0.27 \pm 0.08  \; .
\eea

\subsection{CP asymmetry in flavor specific B decays}
\label{subsec:ASL}
\begin{figure}
\begin{center}
\includegraphics[width=0.48 \linewidth]{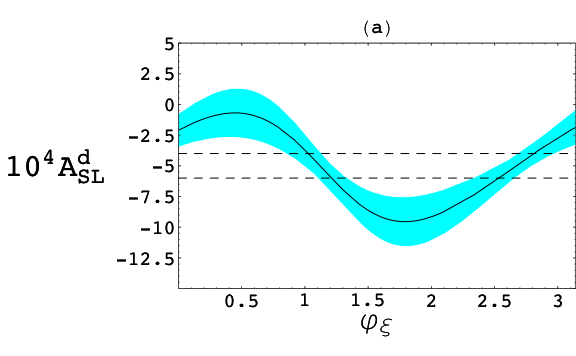}
\raisebox{0.1cm}{\includegraphics[width=0.46 \linewidth]{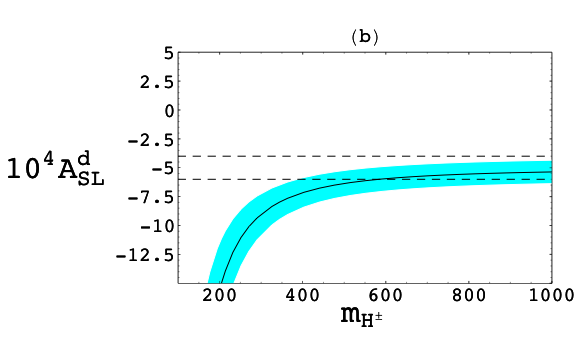}}
\includegraphics[width=0.48 \linewidth]{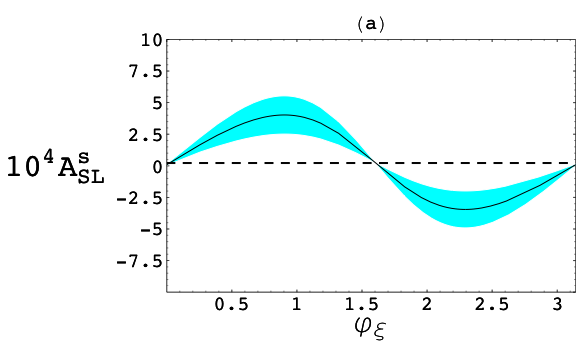}
\raisebox{0.1cm}{\includegraphics[width=0.46 \linewidth]{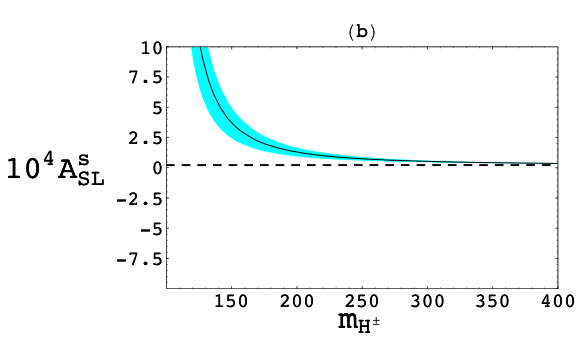}}
\end{center}
\vskip -0.7cm
\caption{
$\varphi_\xi$ and $m_{H^\pm}$ dependence of $A_{SL}^{(d,s)}$ for
$|\xi|=1$ and $\tan\beta_H=50$. The dashed lines are the 1$\sigma$ SM
expectation. The blue band is the theoretical error.
}
\label{fig:asl}
\end{figure}

The CP asymmetry in flavor specific $B_q$ decays (only the decays $B_q
\to f$ and $\bar B_q \to \bar f$ are allowed) is given by:
\bea
A_{SL}^{(q)} & \equiv &
\frac{\Gamma(\bar B_q (t) \to f) - \Gamma(B_q (t) \to \bar f) }{
\Gamma(\bar B_q (t) \to f) + \Gamma(B_q (t) \to \bar f)
}
 =  {\rm Im} \frac{\Gamma_{12}^{(q)}}{M_{12}^{(q)}} \; .
\eea
From the discussion in Sec.~\ref{subsec:deltagamma} it follows that
the T2HDM effects on $\Gamma_{12}^{(q)}$ are negligible, hence we will
consider only box diagram contributions to $M_{12}^{(q)}$. Adopting
the standard parametrization, $M_{12}^{(q)}/ M_{12,SM}^{(q)} = r_q^2
\exp(2 i \theta_q)$ we get:
\bea
A_{SL}^{(q)} & = & {\rm Im} \left( \frac{\Gamma_{12}^{(q)}}{M_{12}^{(q)}} \right)_{\rm SM}
                   \frac{\cos 2 \theta_q}{r_q^2}
                  -{\rm Re} \left( \frac{\Gamma_{12}^{(q)}}{M_{12}^{(q)}} \right)_{\rm SM}
                   \frac{\sin 2 \theta_q}{r_q^2} \; ,
\eea
where~\cite{Beneke:2003az,Ciuchini:2003ww}
\bea
\left( \Gamma_{12}^{(d)}/M_{12}^{(d)} \right)_{\rm SM} & = &
\Big[ -(40 \pm 16) - i \; (5 \pm 1) \Big] \times 10^{-4} \label{GoMd} \\
\left( \Gamma_{12}^{(s)}/M_{12}^{(s)} \right)_{\rm SM} & = &
\Big[ -(40 \pm 16) + i \; (0.22 \pm 0.04) \Big] \times 10^{-4} \; . \label{GoMs}
\eea
In the notation of Sec.~\ref{subsec:Bmixing}, we have $r_q^2 \; \exp(2
i \theta_q) = F_{tt}^{B_q}/S_0(x_t)$. From
Eqs.~(\ref{GoMd},\ref{GoMs}) we read the SM predictions:
\bea
\left(A_{SL}^{(d)}\right)_{\rm SM} &=& (5 \pm 1) \times 10^{-4} \\
\left(A_{SL}^{(d)}\right)_{\rm SM} &=& (0.22 \pm 0.04) \times 10^{-4} \; .
\eea
Unfortunately the experimental errors on these asymmetries are at
least an order of magnitude larger than the SM expectations~\cite{
Jaffe:2001hz,Nakano:2005jb,Aubert:2006nf,Kuze:1999wy,Barate:2000uk,Grossman:2006ce}:
\bea
\left(A_{SL}^{(d)}\right)_{\rm exp} & = & (11 \pm 55) \times 10^{-4} \\
\left(A_{SL}^{(s)}\right)_{\rm exp} & = & (-80 \pm 110) \times 10^{-4} \;.
\eea
In Fig.~\ref{fig:asl} we show the size of the T2HDM contributions to
these asymmetries for large $\tan\beta_H$.

%%%%%%%%%%%%%%%%%%%%%%%%%%%%%%%%%%%%%%%%%%%%%%%%%%%%%%%%%%%%%%%%%%%%%%%
%%%%%%%%%%%%%%%%%%%%%%% FIGURE: ASL %%%%%%%%%%%%%%%%%%%%%%%%%%%%%%%%%%%
%%%%%%%%%%%%%%%%%%%%%%%%%%%%%%%%%%%%%%%%%%%%%%%%%%%%%%%%%%%%%%%%%%%%%%%

%%%%%%%%%%%%%%%%%%%%%%%%%%%%%%%%%%%%%%%%%%%%%%%%%%%%%%%%%%%%%%%%%%%%%%%
%%%%%%%%%%%%%%%%%%%%%%% FIGURE: nEDM %%%%%%%%%%%%%%%%%%%%%%%%%%%%%%%%%%
%%%%%%%%%%%%%%%%%%%%%%%%%%%%%%%%%%%%%%%%%%%%%%%%%%%%%%%%%%%%%%%%%%%%%%%
\begin{figure}
\begin{center}
\includegraphics[width=0.48 \linewidth]{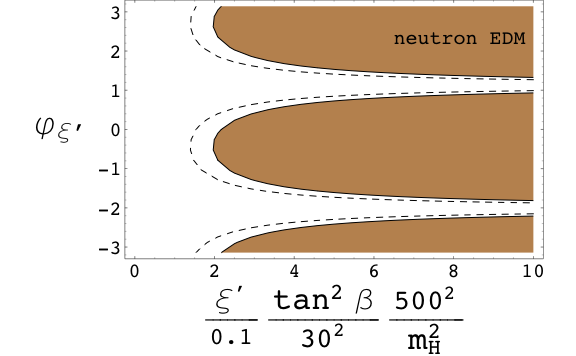}
\end{center}
\vskip -0.7cm
\caption{
Constraint that the neutron EDM puts on the T2HDM parameter space. The
shaded area is excluded at 90\% CL. The solid and dashed lines
correspond to $m_{H^\pm} = (200,1000)$ GeV.
}
\label{fig:nEDM}
\end{figure}
\subsection{Neutron EDM}
\label{subsec:nEDM}
The effective Hamiltonian that encodes charged Higgs contributions to
the neutron EDM is:
\bea
{\cal H}_{\rm eff} & = &
\sum_{q=u,d} C_q \left[
\frac{e}{16\pi^2}  (\bar q_L \sigma^{\mu\nu} q_R) F_{\mu\nu} \; ,
\right]
\eea
where
\bea
C_d & = &
\sum_{i=1}^3
\Bigg\{
\frac{(P_{LR}^H)_{i1}^{} (P_{RL}^H)_{i1}^*}{m_{u_i}}
B \left( m_{u_i}^2 \over m_{H^\pm}^2 \right)+ \cdots
\Bigg\} \;,\\
C_u & = &
\sum_{i=1}^3
\Bigg\{
\frac{(P_{LR}^H)_{1i}^{} (P_{RL}^H)_{1i}^*}{m_{d_i}}
B\left( m_{d_i}^2 \over m_{H^\pm}^2 \right) + \cdots
\Bigg\} \; ,
\eea
and the function $B$ is given in Ref.~\cite{Xiao:2006dq}. The dots
stand for terms that do not contribute to the imaginary part of the
Wilson coefficients. In the chiral quark model, the neutron EDM is
related the valence quark EDM's, $d_u$ and $d_d$,
via~\cite{Ibrahim:1997gj}:
\bea
d_n & = & \frac{1}{3} \left( 4\; d_d -  d_u \right) \; \eta^E \; , \\
d_q & = &  \frac{e}{16\pi^2} \; {\rm Im} (C_q) \; ,
\eea
where $\eta^E \simeq 1.53$ is the QCD correction factor. Approximate
formulae for the up and down quark contributions to the neutron EDM
are (in units of $e \; {\rm cm}$):
\bea
\frac{4}{3} \eta^E d_d & = &
10^{-29} \;
\xi \left(\frac{\tan\beta_H}{30}\right)^2
\left(\frac{500}{m_{H^\pm}}\right)^2 \;
\left[
-(6.5 \pm 0.3) \cos \varphi_{\xi}
-(15.9 \pm 0.7) \sin \varphi_{\xi}
\right] \\
-\frac{1}{3} \eta^E d_u & = &
10^{-26} \;
\frac{\xi^\prime}{0.1} \left(\frac{\tan\beta_H}{30}\right)^2
\left(\frac{500}{m_{H^\pm}}\right)^2 \;
\left[
(3.5 \pm 0.6) \cos \varphi_{\xi^\prime}
- (1.65 \pm 0.3) \sin \varphi_{\xi^\prime}
\right]
\eea
where the uncertainties come from varying $m_{H^\pm}$ in the
$(200-1000)$ GeV range. Taking into account that the 90\% C.L.
experimental upper bound on the neutron EDM is~\cite{Yao:2006px} $6.3
\times 10^{-26} \; e \; {\rm cm}$, it is clear that the T2HDM
parameter space is constrained only for $\xi^\prime \neq 0$. Note that
the huge enhancement in $d_u$ comes from $\Sigma_{13} \propto m_c
\tan\beta_H \xi^\prime/m_W$ that is not suppressed either by $V_{ub}$
or, for large $\tan\beta_H$, by the charm Yukawa.

In Fig.~\ref{fig:nEDM} we show the impact of the present upper bound
on the T2HDM parameter space.

\subsection{CP asymmetries in $B^-\to K^- \pi^0$ and $\bar B^0\to K^- \pi^+$}
The direct CP asymmetries in the decays $B^-\to K^- \pi^0$ and $\bar
B^0\to K^- \pi^+$ can be calculated (albeit with large errors) in the
QCD factorization approach~\cite{Beneke:2003zv}:
\bea
A_{CP} (B^-\to K^- \pi^0)      & = & \left( 7.1^{+1.7+2.0+0.8+9.0}_{-1.8-2.0-0.6-9.7} \right) \% \\
A_{CP} (\bar B^0\to K^- \pi^+) & = & \left( 4.5^{+1.1+2.2+0.5+8.7}_{-1.1-2.5-0.6-9.5} \right) \% \; ,
\eea
where the first error corresponds to variation of the CKM parameters,
the second and third errors refer to uncertainties in the hadronic
parameters used in the calculation and the fourth error reflects
additional uncertainties caused by the breakdown of the factorization
ansatz (that result in endpoint singularities regulated in terms of
two extra complex parameters). Because of a high degree of
correlation, most of these errors cancel when considering the
difference between these two asymmetries. From
Ref.~\cite{Beneke:2003zv}, we see that this difference lies in the
range $[0.5,3.3]$. Using the formulae presented in
Ref.~\cite{Khalil:2005qg} and updated numerical inputs, we find
\bea
\Delta A_{CP} & \equiv & A_{CP} (B^-\to K^- \pi^0) - A_{CP} (\bar B^0\to K^- \pi^+)
 = 2.1 ( 1 \pm 0.5) \; ,
\eea
where the 50\% error reflects the uncertainties studied in
Ref.~\cite{Beneke:2003zv} and the possibility unusually large power
corrections (scenarios S1--S4 of Ref.~\cite{Beneke:2003zv}). This SM
estimate has to be compared to the present experimental
results~\cite{Barberio:2006bi}:
\bea
A_{CP} (B^-\to K^- \pi^0)      & = & \left( 4.7 \pm 2.6 \right) \% \\
A_{CP} (\bar B^0\to K^- \pi^+) & = & \left( -9.7 \pm 1.2 \right) \% \\
\Delta A_{CP} & = & \left( 14.4 \pm 2.9 \right) \% \; .
\eea
The effective Hamiltonian responsible for T2HDM contributions to
$\Delta A_{CP}$ and the relevant matching conditions have been given
in Sec.~\ref{sec:apsiks}.  For completeness we point out that other
approaches to the calculation of $\Delta A_{CP}$ (see, for instance,
Ref.~\cite{Cheng:2004ru}) are consistent with the QCD-factorization
results. Also note that it might be possible to accommodate the present
experimental results in models in which the color--suppressed tree
contribution is unusually enhanced~\cite{Gronau:2006xu}.
%%%%%%%%%%%%%%%%%%%%%%%%%%%%%%%%%%%%%%%%%%%%%%%%%%%%%%%%%%%%%%%%%%%%%%%
%%%%%%%%%%%%%%%%%%%%%%% FIGURE: Delta GAMMA %%%%%%%%%%%%%%%%%%%%%%%%%%%
%%%%%%%%%%%%%%%%%%%%%%%%%%%%%%%%%%%%%%%%%%%%%%%%%%%%%%%%%%%%%%%%%%%%%%%
\begin{figure}
\begin{center}
\includegraphics[width=0.48 \linewidth]{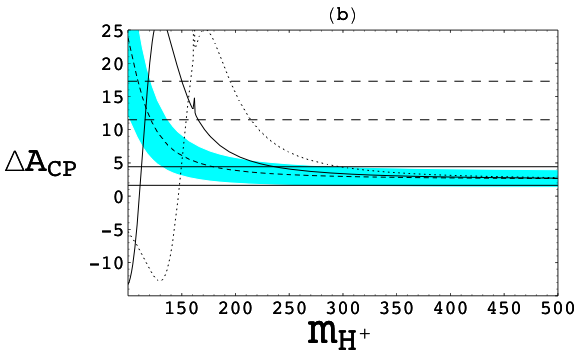}
\raisebox{0.1cm}{\includegraphics[width=0.46 \linewidth]{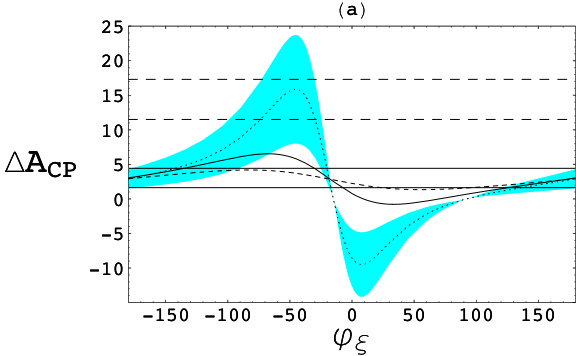}}
\end{center}
\vskip -0.7cm
\caption{
$m_{H^\pm}$ and $\varphi_\xi$ dependence of the T2HDM contributions to
$\Delta A_{CP}$. Solid, dashed and dotted lines correspond to
$(\tan\beta_H,|\xi|)=$ (50,1), (35,1) and (50,2), respectively. The
blue band is the experimental 68\% C.L. allowed region. In plot (a)
and (b) we fix $\varphi_\xi = -50^o$ and $m_{H^\pm}=200\; \gev$,
respectively.}
\label{fig:dacp}
\end{figure}

In Fig.~\ref{fig:dacp} we show the size of T2HDM contributions to
$\Delta A_{CP}$. Unfortunately, for $m_{H^\pm}$ larger than 300 GeV we
do not find any sizable effect.

\section{Observables: the neutral Higgs sector}
\label{sec:neutral}
\subsection{$(g-2)_\mu$}
The anomalous magnetic moment of the muon receive potentially large
contributions at the 2--loop level via the Barr--Zee
mechanism~\cite{Bjorken:1977vt,Barr:1990um}. These diagrams are able
to account for the large discrepancy between the SM prediction and the
experimental measurement only for very light pseudo--scalar mass ($m_A
< 100$ GeV)~\cite{Cheung:2003pw}.

\subsection{$\Delta \rho$}
T2HDM contributions to $\Delta\rho$ depend on both the neutral and
charged Higgs bosons. For any choice of the charged Higgs mass it is
possible to find region of the ($m_A$, $m_H$ and $\alpha_H$) parameter
space for which the corrections to the $\rho$--parameter are in
agreement with the experimental bounds.

\subsection{$Z\to b \bar b$}
Charged and neutral Higgs contributions to the effective $Z-b-\bar b$
coupling affect both the ratio $R_b \equiv \Gamma(Z\to b\bar
b)/\Gamma(Z\to {\rm hadrons})$ and the forward--backward asymmetry
$A_b$. The present experimental results are~\cite{Yao:2006px}:
\bea
R_b & = & 0.21629 \pm 0.00066 \\
A_b & = & 0.901 \pm 0.013 \;,
\eea
where the value for $A_b$ has been obtained by combining the direct
and indirect measurements (from $A_b = 4/3 \; A_{\rm
fb}^{0,b}/A_\ell$). The SM fits for these two observables read:
$R_b^{SM} = 0.21586$ and $A_b^{SM} = 0.935$. In particular note that
$A_b$ shows a deviation of about $2.5\sigma$ from the SM prediction.

The T2HDM contribution to the effective $Z\to b\bar b$ vertex can be
easily extracted from the results of Ref.~\cite{Denner:1991ie}. These
corrections depend on both the charged and neutral Higgs sector of the
T2HDM. Unfortunately we do not find any sizable effect in the portion
of the parameter space allowed by the other constraints.

\section{Summary \& Outlook}
\label{sec:summary}

Thanks to the spectacular performance of the two asymmetric
B-factories, intensive studies in the last few years have demonstrated
that the CP and flavor violation observed in B and K physics is
described by the standard CKM mechanism to an accuracy of about
15\%. A very interesting result, potentially one of the most important
discovery made at the B-factories, is that the time dependent
CP-asymmetries in penguin dominated modes do not seem to agree with
the SM expectations. At the moment these deviations are in the
2.5-3.5$\sigma$ ranges. Since these modes are short-distance
dominated, they are very sensitive to presence of beyond the Standard
Model phases. Taking seriously this pieces of data is suggestive of
sizable contributions from a non-standard phase. For the sake of
completeness we also mentioned several other measurements that display
a significant deviation from the SM, such as difference in the CP
asymmetry between $K^+ \pi^-$ and $K^+ \pi^0$, the (g-2) of the muon
and the forward-backward asymmetry in $Z \to b \bar b$ measured at LEP

As an illustration of a new physics scenario that may account for the
observed deviations from the SM we have presented a detailed study of
the two Higgs doublet model for the top quark. The model is a simple
extension of the SM which in a natural way accounts for the very heavy
top mass. We view it as an interesting low energy effective model that
encompasses some of the important features of an underlying framework
of dynamical electroweak symmetry breaking. Of course, the deviations
seen in B decays and other flavor physics, may also be accountable by
many other extensions of the SM; for example
supersymmetry~\cite{luca}, a fourth family~\cite{ghou}, a Z-penguin
~\cite{gudrun} or warped flavor-dynamics~\cite{aps}. Obviously, the
main features of any extension of the Standard Model that is to
account for the experimental deviations in B-physics and other flavor
physics that are discussed here are that there have to be new
particles in the $\approx$ 300 GeV to $\approx$ few TeV range and
associated with these one needs at least one new CP-odd phase.
Distinguishing between various scenarios or nailing down the precise
structure of some other models responsible for the deviations that we
have discussed will undoubtedly require much more experimental
information. In any case if the hints from the B-factories are really
true, we will certainly witness a truly exciting era in Particle
Physics as the LHC starts its long awaited operations in 2008. It
should also be abundantly clear that infusion of precise information
from low energy flavor measurements will be crucial in interpreting
the findings at the terascale energy.

\section*{Acknowledgments}
This research was supported in part by the U.S. DOE contract
No.DE-AC02-98CH10886(BNL). Research partly supported by the Department
of Energy under Grant DE-AC02-76CH030000. Fermilab is operated by
Fermi Research Alliance, LLC under Contract No. DE-AC02-07CH11359 with
the United States Department of Energy.

\end{document}